\documentclass[%
 reprint,
 superscriptaddress,
 amsmath,amssymb,
 aps,
floatfix,
]{revtex4-2}

\usepackage{graphicx}
\usepackage{dcolumn}
\usepackage{bm}
\usepackage{xcolor}
\usepackage{multirow}
\usepackage{upgreek}

\newcommand{\ket}[1]{\left\vert{#1}\right\rangle}

\usepackage{textcomp}
\usepackage{xcite}
\usepackage{verbatim}
\usepackage[colorinlistoftodos]{todonotes}

\usepackage{xr-hyper}
\usepackage{hyperref}
\makeatletter
\newcommand*{\addFileDependency}[1]{
  \typeout{(#1)}
  \@addtofilelist{#1}
  \IfFileExists{#1}{}{\typeout{No file #1.}}
}
\makeatother
\newcommand*{\myexternaldocument}[1]{
    \externaldocument{#1}
    \addFileDependency{#1.tex}
    \addFileDependency{#1.aux}
}
\myexternaldocument{supplement}

\begin{document}

\preprint{APS/123-QED}

\title{Fast high-fidelity single-shot readout of spins in silicon using a single-electron box}

\author{G. A. Oakes}
 \email{These authors contributed equally to this work}
 \affiliation{Cavendish Laboratory, University of Cambridge, J.J. Thomson Avenue, Cambridge CB3 0HE, United Kingdom}
 \affiliation{Quantum Motion, 9 Sterling Way, London N7 9HJ, United Kingdom}
\author{V.N. Ciriano-Tejel}%
 \email{These authors contributed equally to this work}
 \affiliation{Quantum Motion, 9 Sterling Way, London N7 9HJ, United Kingdom}
 \affiliation{London Centre for Nanotechnology, University College London, London WC1H 0AH, United Kingdom}
\author{D. Wise}%
 \affiliation{Quantum Motion, 9 Sterling Way, London N7 9HJ, United Kingdom}\affiliation{London Centre for Nanotechnology, University College London, London WC1H 0AH, United Kingdom}
\author{M. A. Fogarty}%
 \affiliation{Quantum Motion, 9 Sterling Way, London N7 9HJ, United Kingdom}
 \affiliation{London Centre for Nanotechnology, University College London, London WC1H 0AH, United Kingdom}
 \author{T. Lundberg}%
 \affiliation{Cavendish Laboratory, University of Cambridge, J.J. Thomson Avenue, Cambridge CB3 0HE, United Kingdom}
 \affiliation{Hitachi Cambridge Laboratory, J.J. Thomson Avenue, Cambridge CB3 0HE, United Kingdom}
\author{C. Lain\'{e}}%
 \affiliation{Quantum Motion, 9 Sterling Way, London N7 9HJ, United Kingdom}
\affiliation{London Centre for Nanotechnology, University College London, London WC1H 0AH, United Kingdom}
 \author{S. Schaal}%
 \affiliation{Quantum Motion, 9 Sterling Way, London N7 9HJ, United Kingdom}
\affiliation{London Centre for Nanotechnology, University College London, London WC1H 0AH, United Kingdom}
 \author{F.~Martins}%
 \affiliation{Hitachi Cambridge Laboratory, J.J. Thomson Avenue, Cambridge CB3 0HE, United Kingdom}
\author{D. J. Ibberson}%
 \affiliation{Quantum Engineering Technology Labs, University of Bristol, Tyndall Avenue, Bristol BS8 1FD, United Kingdom}
 \affiliation{Hitachi Cambridge Laboratory, J.J. Thomson Avenue, Cambridge CB3 0HE, United Kingdom}
\author{L. Hutin}%
 \affiliation{CEA, LETI, Minatec Campus, F-38054 Grenoble, France}
\author{B. Bertrand}%
 \affiliation{CEA, LETI, Minatec Campus, F-38054 Grenoble, France}
\author{N. Stelmashenko}%
 \affiliation{Department of Materials Science and Metallurgy, University of Cambridge,
27 Charles Babbage Road, Cambridge CB3 0FS, United Kingdom}
\author{J. A. W. Robinson}%
 \affiliation{Department of Materials Science and Metallurgy, University of Cambridge,
27 Charles Babbage Road, Cambridge CB3 0FS, United Kingdom}
\author{L. Ibberson}%
 \affiliation{Hitachi Cambridge Laboratory, J.J. Thomson Avenue, Cambridge CB3 0HE, United Kingdom}
\author{A.~Hashim}%
 \affiliation{Quantum Nanoelectronics Laboratory, Dept. of Physics, Univ. of California, Berkeley CA 94720, USA}
\author{I. Siddiqi}%
 \affiliation{Quantum Nanoelectronics Laboratory, Dept. of Physics, Univ. of California, Berkeley CA 94720, USA}
\author{A. Lee}%
 \affiliation{Cavendish Laboratory, University of Cambridge, J.J. Thomson Avenue, Cambridge CB3 0HE, United Kingdom}
\author{M. Vinet}%
 \affiliation{CEA, LETI, Minatec Campus, F-38054 Grenoble, France}
\author{C. G. Smith}
 \affiliation{Cavendish Laboratory, University of Cambridge, J.J. Thomson Avenue, Cambridge CB3 0HE, United Kingdom}
 \affiliation{Hitachi Cambridge Laboratory, J.J. Thomson Avenue, Cambridge CB3 0HE, United Kingdom}
\author{J.J.L. Morton}
 \email{john@quantummotion.tech}
 \affiliation{Quantum Motion, 9 Sterling Way, London N7 9HJ, United Kingdom}
\affiliation{London Centre for Nanotechnology, University College London, London WC1H 0AH, United Kingdom}
\author{M. F. Gonzalez-Zalba}
 \email{fernando@quantummotion.tech}
 \affiliation{Quantum Motion, 9 Sterling Way, London N7 9HJ, United Kingdom}

\date{\today}

\begin{abstract}
 
Three key metrics for readout systems in quantum processors are measurement speed, fidelity and footprint. Fast high-fidelity readout enables mid-circuit measurements, a necessary feature for many dynamic algorithms and quantum error correction, while a small footprint facilitates the design of scalable, highly-connected architectures with the associated increase in computing performance. 
Here, we present two complementary demonstrations of fast high-fidelity single-shot readout of spins in silicon quantum dots using a compact, dispersive charge sensor: \emph{a radio-frequency single-electron box}. The sensor, despite requiring fewer electrodes than conventional detectors, performs at the state-of-the-art achieving spin read-out fidelity of 99.2\% in less than 6~$\upmu$s. We demonstrate that low-loss high-impedance resonators, highly coupled to the sensing dot, in conjunction with Josephson parametric amplification are instrumental in achieving optimal performance. We quantify the benefit of Pauli spin blockade over spin-dependent tunneling to a reservoir, as the spin-to-charge conversion mechanism in these readout schemes.
Our results place dispersive charge sensing at the forefront of readout methodologies for scalable semiconductor spin-based quantum processors. 

\end{abstract}

\maketitle

Electron spin qubits in silicon are consolidating their position as a leading candidate to build scalable high-fidelity quantum processors. Several recent demonstrations have shown single- and two-qubit gate fidelities exceeding the requirements for fault tolerant thresholds in the same device~\cite{xue2021a, Noiri2021, Mills2021}.
Combined with the dense scaling potential~\cite{Veldhorst2017, Boter2021}, advanced manufacturing~\cite{Maurand2016, Zwerver2021} and prospects for integration with cryogenic classical electronics~\cite{Ruffino2021a}, these results present a promising future for spin-based qubits in silicon.

For universal quantum computing, the technology will require fast high-fidelity readout on a timescale which is short compared to the qubit coherence time to allow for error correction codes to be implemented. Even for noisy intermediate-scale quantum (NISQ) processors, fast measurement remains advantageous to avoid readout becoming the bottleneck in circuit run-time and to enable mid-circuit measurements for error mitigation\cite{McArdle2019, Botelho2021} and gate teleportation~\cite{Wan2019}.

Fast spin readout has been shown using single-electron transistors (SETs) with spin-readout fidelities as high as 99.9\% in 6 $\mu$s~\cite{Curry2019} and 99\% in 1.6~$\mu$s in radiofrequency (rf) mode~\cite{Connors2020}. However, an SET requires at least three electrodes and two charge reservoirs significantly limiting their placement within dense qubit arrays.

Dispersive readout methods, based on detecting alternating single-electron currents, offer the benefit that only two electrodes are needed to either sense the system \textit{in-situ} using Pauli spin blockade (PSB)~\cite{Petersson2010, Colless2013, GonzalezZalba2015} or to create a dispersive charge sensor, i.e. a single-electron box (SEB), to read the target qubit~\cite{House2016, Urdampilleta2019, Chanrion2020, Ansaloni2020, CirianoTejel2021}. These methods require zero or one charge reservoirs, respectively, strongly enhancing their capacity to be introduced within qubit arrays. The SEB, however, offers the key technological advantage it can detect electronic transitions occurring at rates much lower than the probing rf frequency, a common necessity in the few-electron regime where qubits are operated. However, the performance of SEBs in silicon has remained non-competitive with respect to SETs (99.2\% in 100~$\mu$s~\cite{Borjans2021}), raising the question of whether fast and compact high-fidelity readout could be possible.

Here, we present two independent demonstrations showing that the compact SEB electrometer can achieve fast and accurate spin readout in silicon, reaching for example, 99.2\% fidelity in just 5.6 $\mu$s. Our work spans four key aspects in designing optimal dispersive charge sensors: (i) the SEB design, (ii) the readout resonator design, (iii) the physical mechanism for spin-to-charge conversion and (iv) the amplification chain. We present SEBs with large gate couplings (i) coupled to low-loss high-impedance rf resonators (ii). We show the benefits of Pauli spin blockade readout over spin-dependent tunneling to a reservoir (iii) and demonstrate the role of parametric amplification in improving dispersive readout fidelity (iv). These results demonstrate a route to combining high-fidelity readout of semiconductor-based qubits with the demands of a compact and scaleable architecture.

\section{Optimising readout sensitivity}\label{sec:intro-theory}

Dispersive charge sensing using an SEB works on the principle of some change $\Delta C_{\rm D}$ in the quantum capacitance of the SEB arising from a change in its local electrostatic environment. This capacitance shift is detected via a change in the reflected power, $\Delta P_\mathrm{rf}$, of an rf signal from a resonator whose frequency depends on $C_{\rm D}$. The sensitivity of the SEB can be characterized by the ratio of the signal and noise powers, $\mathrm{SNR}=\Delta P_\mathrm{rf}/P_\mathrm{n}$, where $P_\mathrm{n}$ is the noise power. In order to maximise the SNR, the available strategies are to minimise the noise power; or to maximise the signal power. As we shall see later, the signal power is enhanced by i) increasing the gate lever arm of the SEB, $\alpha$ (defined as the ratio of gate capacitance of the SEB and its total capacitance) and its capacitive coupling to the spin system to be sensed; and by (ii) optimizing the resonator design which involves a well matched, low loss, high impedance resonator, operating at high frequency but kept below the tunnel rate of the SEB.
More details on the dependence of the SNR on each of these parameters are given below (see Eq.~\eqref{eq:tau_m} and \S I).
Here, we present different approaches to increasing the SNR by optimisation of these parameters, and compare two strategies for converting the spin degree of freedom of the qubit into a charge event which can be detected by the SEB. We begin by using spin-dependent tunneling, while minimising the noise power $P_\mathrm{n}$ through use of a Josephson Parametric Amplifier (JPA)~\cite{Vijay2009}. 

\section{Spin-dependent tunneling}\label{sec:Elzerman}

\begin{figure}
\centering
\includegraphics[width=1\linewidth]{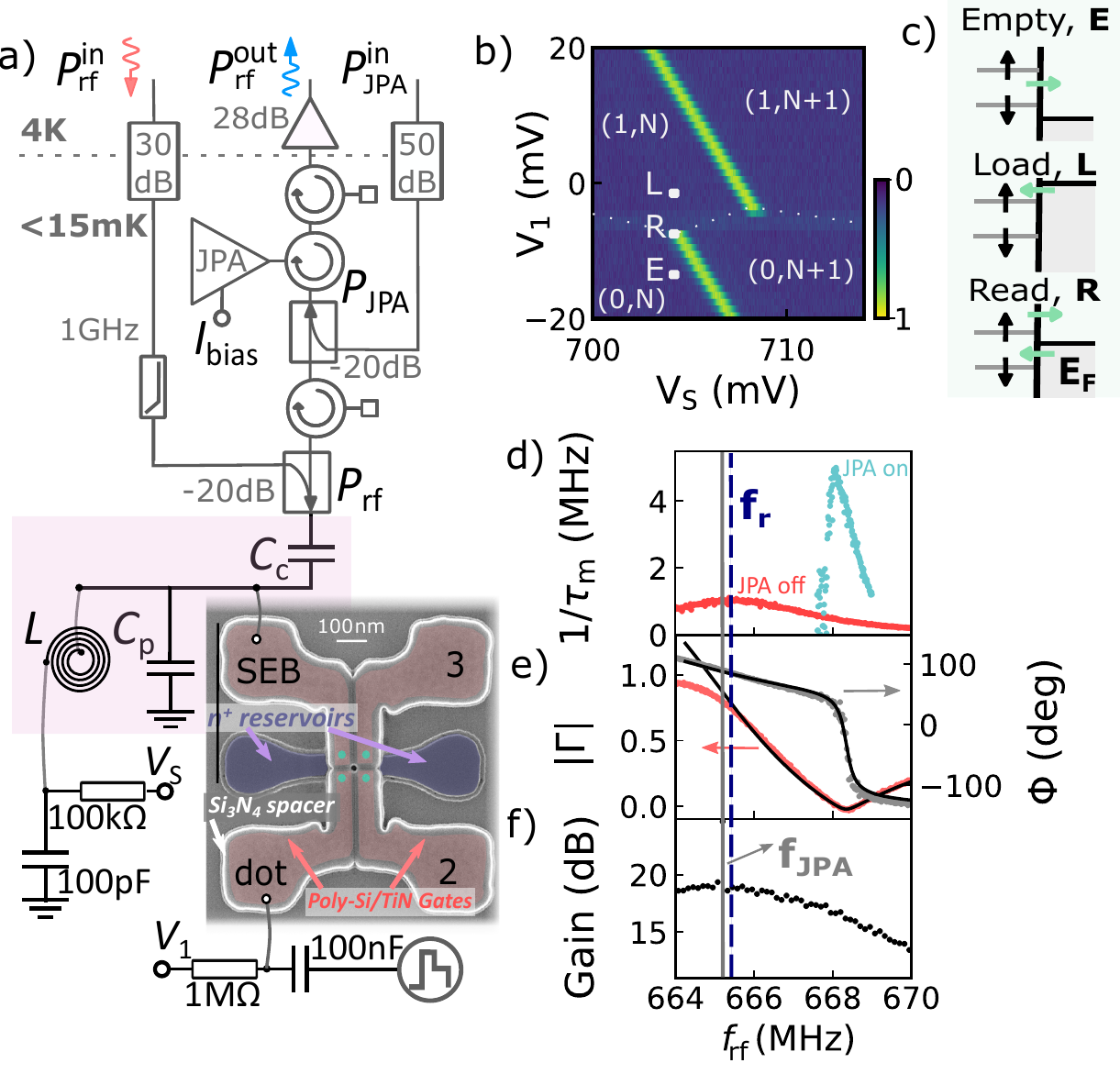}
\caption{Device and measurement setup. a) False-colour scanning electron micrograph of a silicon nanowire transistor with two pairs of split gates (red). The green dots indicate the location of the QDs under each gate. The blue regions are electron reservoirs. The SEB is connected to a lumped-element resonator
and the fast pulses for spin-dependent readout are applied to the dot facing it through a bias tee. 
b) Normalised rf-response of the stability diagram of the SEB and the dot. The SEB potential shifts when the first electron is added to the dot, changing the rf-response from a maximum to some minimum, background level. c) The dot is first emptied [E] so an electron with a random spin polarisation enters the dot at the load stage [L]. Finally the spin polarisation is measured [R] by placing the reservoir in between the $\ket{\uparrow}$ and $\ket{\downarrow}$ states.
d) The inverse of the minimum integration time $1/\tau_\mathrm{m}$ for a SNR=1 is shown as a function of $f_\mathrm{rf}$ without and with the JPA (red and blue respectively). The black dashed line indicates the natural frequency of the resonator, $f_\mathrm{res}$, which optimises sensitivity without the JPA. e) The reflection coefficient (magnitude $|\Gamma|$ (red) and phase $\Phi$ (grey)) at $B=2$~T, shows a minimum at the frequency which optimises sensitivity with the JPA. 
f) The JPA gain decreases as $f_\mathrm{rf}$ moves away from the JPA pumping frequency, $f_\mathrm{JPA}$, solid grey line, as determined by the JPA bandwidth of 19.2~MHz. } 
    	\label{fig:Fig1}
        \end{figure}
        
The device in this section is an etched silicon nanowire double split-gate transistor that, at low temperatures, can be used to form quantum dots (QDs) at the upper edges of the nanowire, one under each of the four gates (See Fig.~\ref{fig:Fig1}a and Methods for device dimensions). One of the gates is connected to a lumped-element resonator to form an SEB that is used as a charge sensor to measure the qubit --- here, the spin of an electron confined
under the gate opposite to the SEB ---
using spin-dependent tunneling. The other two gates are not used. 
Changes in the SEB quantum capacitance, $C_\mathrm{D}$ are detected via rf reflectometry~\cite{Vigneau2022}, where a tone of frequency $f_\mathrm{rf}$ is delivered through an attenuated line to the lumped resonator (see Fig.~\ref{fig:Fig1}a) capacitively coupled to the transmission line ($C_\mathrm{c}=50$ fF). The resonator, of natural frequency $f_\mathrm{res}$=665.5~MHz at 2~T, consists of a spiral NbN inductor ($L=124$~nH) and the parasitic capacitance ($C_\mathrm{p}=411$ fF) in parallel with the SEB and presents an internal Q-factor, $Q_0=270$, a high-impedance $Z_\text{r}=519~\Omega$ and a coupling coefficient $\beta=2.5$ (see \S II).

The SEB, with an $\alpha=0.35$, is \emph{strongly sensitive} to the `qubit dot' (as shown in Fig.~\ref{fig:Fig1}b), meaning 
that the addition of one electron to the QD causes a change in the quantum capacitance of the SEB which shifts its rf-response by more than one linewidth. This shift is used to perform spin readout using the three-level pulse sequence illustrated in Fig.~\ref{fig:Fig1}b-c. 

The noise of the system (conveniently characterised by the equivalent noise temperature $T_\text{n}$) contains contributions from the system itself $T_\text{sys}$ (SEB and resonator) as well as the amplification chain (see Fig.~\ref{fig:Fig1}a). In rf measurements of semiconductor QDs, a high-mobility electron transistor (HEMT) is typically used as the first amplifier, limiting the noise temperature to a few Kelvin. Here we add an amplifier with lower noise temperature, a JPA, with gain $G_\mathrm{JPA}$, reducing $T_\mathrm{n}$ accordingly:
\begin{equation}
    T_\mathrm{\rm n}=T_\mathrm{\rm sys}+T_\mathrm{\rm JPA}+\frac{T_\mathrm{\rm HEMT}}{G_\mathrm{\rm JPA}}.
\end{equation}
where $T_\mathrm{HEMT(JPA)}$ is the noise temperature of the HEMT(JPA). We note the use of a JPA is only advantageous if the noise is dominated by the HEMT instead of the system. This is not the case for rf-SETs where shot noise may be comparable or in excess of that of the HEMT~\cite{Keith2019}. For SEBs, if the tunneling rate $\gamma$ between SEB and reservoir is greater than $f_\mathrm{rf}$, electrons tunnel adiabatically, and the Sisyphus noise vanishes~\cite{GonzalezZalba2015} leaving predominantly the noise contribution of the HEMT (see \S III for a description of the JPA).

In this device, the tunneling rate $\gamma$ between SEB and reservoir is $74\pm12$~GHz (see \S IV), well in excess of $f_\mathrm{rf}\sim0.7$~GHz, such that we are in the regime of negligible Sisyphus dissipation in the SEB. This is confirmed by noise temperature measurements yielding $T_\mathrm{n}=2.5^{+1.4}_{-0.9}$~K~\cite{Schaal2020} with the JPA off, reducing to $T_\mathrm{n}=0.25^{+0.14}_{-0.09}$~K with the JPA. The fact that the latter falls below typical shot noise levels~\cite{Keith2019} demonstrates one of the major advantages of SEBs over SETs.
This tenfold reduction of noise temperature with the JPA leads to a corresponding reduction in $\tau_\mathrm{m}$, the minimum integration time to resolve a charge event with a SNR=1~\cite{Stehlik2015}. When measuring at $f_{\rm rf}= 668$~MHz, where the reflection coefficient of the resonant circuit is at a minimum, we find $\tau_\mathrm{m}=2$~$\mu$s and $\tau_\mathrm{m}=200$~ns, for JPA off and on, respectively.
Operating at this point of minimum reflected power is necessary to avoid driving the JPA beyond its 1 dB compression point, $P_\text{1dB}=-116$~dBm. However, as can be seen in Fig.~\ref{fig:Fig1}d, $\tau_\mathrm{m}$ with the JPA off can be decreased by approximately a factor of two by adjusting the drive frequency $f_{\rm rf}$ to match $f_{\rm res}$, which differs from the point of minimum reflected power in the total circuit. The overall achievable reduction in $\tau_\mathrm{m}$ achieved using the JPA is therefore a factor of 4.5 (see \S V for more details). These results emphasise the importance of a well-matched and high-Q resonator 
to minimise the reflected power
to avoid saturating the JPA~\cite{Schaal2020}.
The limit in readout bandwidth is set by the difference $\Delta f$ between the JPA pump frequency ($f_\mathrm{JPA}=665.2$~MHz) and $f_\mathrm{rf}$, while the JPA gain falls as this difference increases (see Fig.~\ref{fig:Fig1}f). We select $\Delta f=2.9$~MHz for which $G_\mathrm{JPA}=17$~dB.
     
\subsection{Spin readout and fidelity}\label{sec:readout_elzerman}   

\begin{figure*}
\centering
\includegraphics[width=1\textwidth]{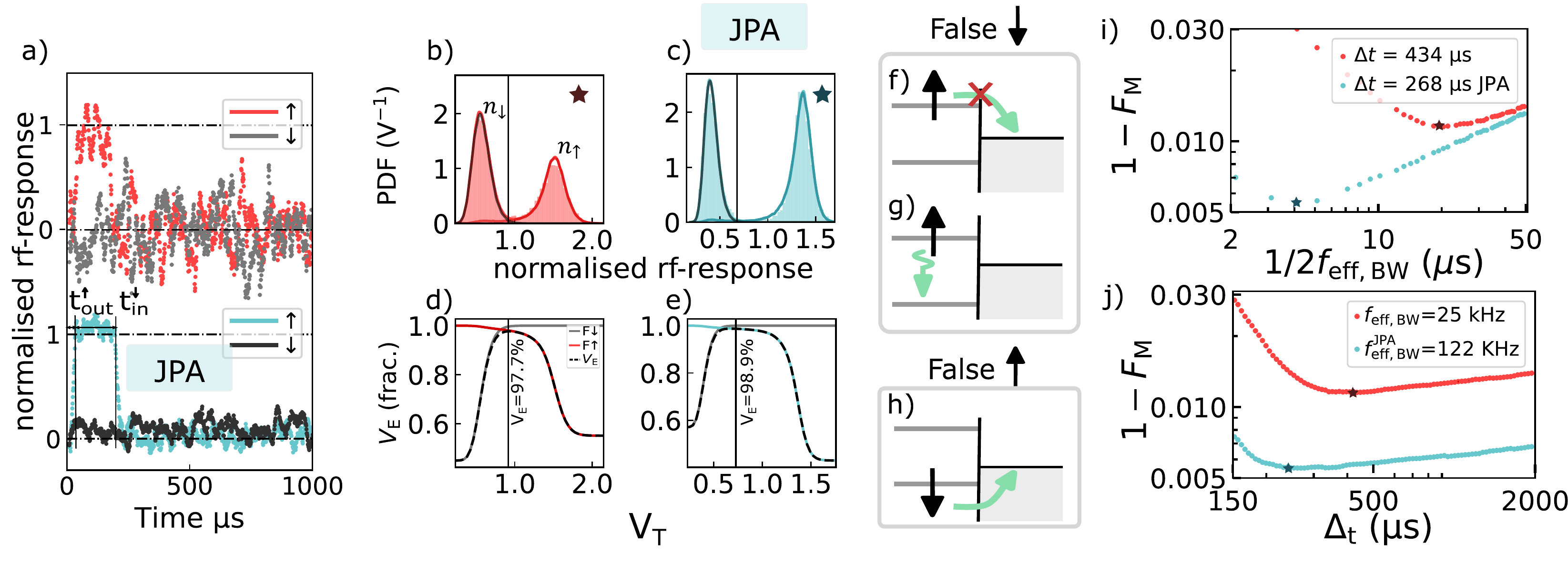}
\caption{ Spin readout fidelity. a) Top: spin $\ket{\uparrow}$ (red) and $\ket{\downarrow}$ (grey) traces taken without the JPA. The rf-response is normalised so it is 0 when the dot is occupied and 1 when it's empty.
The bottom panel shows spin $\ket{\uparrow}$ (blue) and spin $\ket{\downarrow}$ (black) traces taken with a JPA. b) Histogram of the maximum values of the normalised rf-response from 10,000 single-shot measured data traces taken without a JPA. The red line is the simulated histogram created using the parameters $A= 0.50$, $\mathrm{\Gamma_\mathrm{s}=}1$~MHz, ${t_\mathrm{out}^\uparrow}=53$~$\mathrm{\mu s}$, ${t_\mathrm{in}^\downarrow}=440$~$\mathrm{\mu s}$ and Gaussian noise with standard deviation $\sigma_{high}=1.09$ for the top of the blip and and $\sigma_{low}=1.03$ for the background. Both the readout bandwidth and readout time used to create this histogram corresponds with the optimal $\Delta t=434$~$\mathrm{\mu s}$ and readout bandwidth $f_\mathrm{eff,BW}=25$~kHz. c) Same as b) for measurement taken with a JPA. In this case, the parameters used for the simulation are $A_\mathrm{JPA}=0.46$, ${\Gamma_\mathrm{s,JPA}}=1$~MHz, ${t_\mathrm{out, JPA}^\uparrow}=31$~$\mathrm{\mu s}$, ${t_\mathrm{in, JPA}^\downarrow}=186$ ~$\mathrm{\mu s}$ and Gaussian noise with standard deviation $\sigma_\mathrm{high,JPA}=0.38$ for the top of the blip and $\sigma_\mathrm{low,JPA}=0.36$ for the background. The post-processing parameters are chosen to be the ones that maximises the visibility ($\Delta t=268$~$\mathrm{\mu s}$ and $f^\mathrm{JPA}_\mathrm{eff,BW}=122$~kHz). d) Electrical visibility, $V_\mathrm{E}$, as a function of the threshold voltage used to discriminate between spin down and up.  e) Same for traces obtained using a JPA.  f, g, h) Spin mapping errors due to a short readout time with respect to ${t_\mathrm{out}^\uparrow}$ (f), relaxation processes (g) or thermal excitations (f). i) Measurement infidelity ($1-F_\mathrm{M}$) taken with (blue) and without (red) a JPA as a function of the integration time, which is equal to $1/2f_\mathrm{eff,BW}$. j) Measurement infidelity versus measurement time, $\Delta t$. The stars mark the optimal integration times and measurement times.
} 
\label{fig:Fig2}
\end{figure*}

To measure the spin of the electron in the QD, we apply a magnetic field of $B=2$~T to produce a Zeeman splitting $E_\mathrm{z}=g\mu_\mathrm{B}B$ larger than the thermal broadening of the reservoir into which the electron tunnel. Here, $T_\text{e}=137\pm 18$~mK is the electron temperature (see \S VI), $\mu_B$ is the Bohr magneton and $g\approx 2$ is the electron g-factor. To measure the spin orientation, we apply a 3-level voltage pulse to the QD gate (see inset of Fig.~\ref{fig:Fig1}b). First, the QD is emptied so an electron with a random spin polarisation can be loaded from the reservoir. Then, at the readout stage, the reservoir Fermi energy, $E_\mathrm{F}$, lies in between the spin $\ket{\uparrow}$ and $\ket{\downarrow}$ states, so a spin $\ket{\uparrow}$ electron can tunnel out from the dot to the reservoir and be subsequently replaced by a spin $\ket{\downarrow}$ electron, whereas a spin $\ket{\downarrow}$ electron remains in the QD~\cite{Elzerman2004}. During the readout stage, the system is tuned at the position marked `R' in Fig.~\ref{fig:Fig1}b, where the SEB rf-response is strongly dependent on the QD electronic occupation. This way, a readout trace from a spin $\ket{\downarrow}$ state is a constant noisy background (the grey and black traces in Fig.~\ref{fig:Fig2}a). On the other hand, a spin $\ket{\uparrow}$ readout trace is characterised by a top hat shape that starts when the spin $\ket{\uparrow}$ electron leaves the dot  (${t_\mathrm{out}^\uparrow}$), and lasts until a spin $\ket{\downarrow}$ electron tunnels back into the dot  (${t_\mathrm{in}^\downarrow}$). Spin $\ket{\uparrow}$ single-shot traces taken without the JPA are displayed in red in Fig.~\ref{fig:Fig2}a, whereas the ones using a JPA in blue, show a noticeable $\times 4.5$ SNR improvement. In both cases, the traces are taken with a sample rate of $\mathrm{\Gamma_\mathrm{s}}=1$~MHz and an readout bandwidth of $f_\mathrm{eff,BW}=25$~kHz (See \S VII for more information about the experiment bandwidth).

We identify the spin polarisation of a given trace by setting a threshold in the rf-response that is compared against the trace's maximum. If the threshold is exceeded, the trace is labeled as a spin $\ket{\uparrow}$ and if not, as $\ket{\downarrow}$. The trace's maxima follow a bimodal probability distribution, as in Fig.~\ref{fig:Fig2}b and c, with one peak corresponding to spin $\ket{\uparrow}$ traces and the other to $\ket{\downarrow}$ traces. To determine the readout fidelity, we model the histograms as $N_\mathrm{tot}(V_\mathrm{rf})=N_\text{tot}[n_\uparrow(V_\mathrm{rf})+n_\downarrow(V_\mathrm{rf})]V_\mathrm{bin}$, where $n_{\uparrow(\downarrow)}$ is the probability density of the maxima of spin $\ket{\uparrow}$ ($\ket{\downarrow}$) traces, $V_\mathrm{rf}$ is the normalised rf-response, $N_\mathrm{tot}$ is the total number of traces and $V_\text{bin}$ is the width of the rf-response bins~\cite{Barthel2010}. The fidelity of correctly labelling an individual readout trace, $F\mathrm{_E^{\uparrow(\downarrow)}}$, can be calculated as:

\begin{equation}
    F\mathrm{_E^\downarrow}=1-\int_{V_\mathrm{T}}^{\infty}  n_\downarrow(V_\mathrm{rf}) \, dV_\mathrm{rf} \,\,\,\,\,\,\,\,\,\,\,  F\mathrm{_E^\uparrow}=1-\int_{-\infty}^{V_\mathrm{T}}  n_\uparrow(V_\mathrm{rf}) \, dV_\mathrm{rf},
\end{equation}
where the integral of $n_{\downarrow(\uparrow)}$ from $V_\mathrm{T}(-\infty)$ to $\infty(V_\mathrm{T})$ is the cumulative probability of having labeled spin $\ket{\downarrow}(\ket{\uparrow})$ trace wrongly~\cite{Barthel2010}.

The experimental data results in the bimodal distribution as a whole. However, to obtain $n_{\uparrow}$ and $n_{\downarrow}$ separately, we numerically generate 100,000 readout traces, where each trace is assigned a spin polarisation with probability $A$ of being spin $\ket{\downarrow}$ and $1-A$ of being spin $\ket{\uparrow}$. Readout traces are completely determined by few experimental parameters that can be extracted from the data traces: the sample rate $\Gamma_\text{s}$, the readout bandwidth, $f_\mathrm{eff,BW}$, the sensor SNR, the tunneling times
$t_\mathrm{out}^\uparrow$, and
$t_\mathrm{in}^\downarrow$,  and the readout time, $\Delta t$ (See \S VIII for a description of the parameter extraction).
We fit the simulated histogram to 10,000 experimental shots using least squares regression, see Fig.~\ref{fig:Fig2}b and c. In Fig.~\ref{fig:Fig2}c, $n_{\uparrow}$ (solid black curve) and $n_{\downarrow}$ (solid blue curve) are comparatively narrower due to the reduced noise enabled by the JPA.
As shown in Fig.~\ref{fig:Fig2}d and e, the electrical visibility, $V_\mathrm{E}= 1-F\mathrm{_E^\uparrow}-F\mathrm{_E^\downarrow}$, depends on the selected threshold voltage, $V_\mathrm{T}$. We obtain $V\mathrm{_E}= 97.7\%$ without a JPA and of $V\mathrm{_E}^\mathrm{JPA}= 98.9\%$ using a JPA. 

Depending on the readout time, $\Delta t$, spin mapping errors can diminish the readout fidelity. If $\Delta t$ is of the order or smaller than $t_{out}^\uparrow$, spin $\ket{\uparrow}$ electrons won't leave the QD during the readout time , leading to a false spin $\ket{\downarrow}$ measurement (See Fig.~\ref{fig:Fig2}f). On the other hand, if $\Delta t$ is increased, a spin $\ket{\uparrow}$ may relax to the ground state before leaving the QD, resulting in a false spin $\ket{\downarrow}$ (See Fig.~\ref{fig:Fig2} g), or a spin $\ket{\downarrow}$ could be thermally excited out of the QD, leading to a false spin $\ket{\uparrow}$ (See Fig.~\ref{fig:Fig2} h). The spin readout fidelity, $F_\mathrm{M}$ is the product of the electrical fidelity, $F_\mathrm{E}$, which determines the probability to label correctly a given readout trace (as discussed earlier) and the spin-to-charge fidelity, $F_\mathrm{STC}$, setting the probability that a spin state generates the trace that it's expected to. In our case, $F_\mathrm{E}$ also
includes the false negatives derived from a slow $ t_{out}^\uparrow$, since the simulated traces have a finite length, $\Delta t$ 
(See \S IX for a full description).

Having taken spin mapping errors in consideration, we investigate the dependence of $F_\text{M}$ on
$\Delta t$, and $f_\mathrm{eff,BW}$ (See \S X for a full discussion). Figure~\ref{fig:Fig2}i shows how decreasing $f_\mathrm{eff,BW}$ leads to an improved fidelity as the noise is reduced, up to a point in which the additional filtering deforms the spin $\ket{\uparrow}$ top hat, smoothing the edges and reducing its maximum value. The optimal readout bandwidth is different for measurements taken with and without the JPA not only because of the SNR improvement, but also because of the different tunneling rates in each data set, caused by a shift of the readout point and the 1d nature of the reservoir (See \S VI).

Further, the fidelity improves as $\Delta t$ is increased since more $\ket{\uparrow}$ are captured, see Fig.~\ref{fig:Fig2}j. However, beyond an optimal value, the fidelity worsens because of the additional opportunities for the background noise to surpass the threshold. Spin mapping errors due to thermal excitation and relaxation are negligible due to their large time constant, being the relaxation time $T_1=5.2$~ms and the time constant for a thermal excitation ${t_\mathrm{out}^\downarrow}= 309$~s and 70~s without and with the JPA, respectively.

We obtain a maximum spin readout fidelity $F_\mathrm{M}=98.85\%$ without the JPA at $\Delta t=434$~$\mu$s and $f_\mathrm{eff,BW}=25$~kHz and $F_\mathrm{M,JPA}=99.45\%$ for measurements obtained with a JPA using $\Delta t_\mathrm{JPA}=268$~$\mu$s and $f^\mathrm{JPA}_\mathrm{eff,BW}=122$~kHz. We note $F_\mathrm{M, JPA}=99\%$ is already achieved at $\Delta t_\mathrm{JPA}=131$~$\mu$s. We further explore machine learning-based approaches to improve readout fidelity~\cite{Struck2021, Matsumoto2020}. Here, by using Neural Networks, we report an increased fidelity of $F_\mathrm{M}=99.1\%$ in $\Delta t=500$~$\mu$s, and $F_\mathrm{M,JPA}=99.54\%$ in $\Delta t=250$~$\mu$s without and with a JPA respectively (See \S XI for more information).

Finally, we analyze the impact of $t_\mathrm{out}^\uparrow$ on the readout fidelity. During this period, spin $\ket{\uparrow}$ and $\ket{\downarrow}$ cannot be differentiated adding idle time to $\Delta t$. We simulate readout traces with asymmetric tunnel rates $t_\mathrm{out}^\uparrow \ll 1$~$\mu$s and $t_\mathrm{in}^\downarrow=228$~$\mu$s and obtain $F_\mathrm{M}=99.3\%$ in just 4~$\mu$s (See \S XII for a full discussion). The result show that the SEB could assign a spin label in much shorter timescales with equivalent fidelity if a spin-to-charge conversion mechanism with this tunneling characteristics could be used.

\section{Pauli Spin Blockade}\label{sec:PSB}
\begin{figure}[ht]
    \centering
    \includegraphics[width=\linewidth]{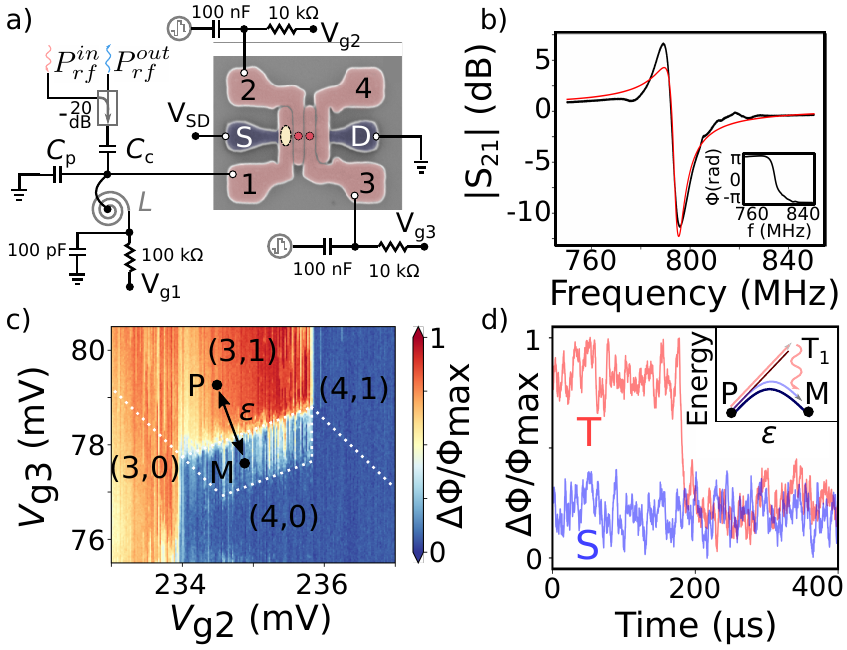}
    \caption{Experimental setup. a) False coloured scanning electron micrograph of a silicon nanowire transistor with wrap around gates, under which QDs form. The QD under gate one is operated in the many electron regime (yellow ellipse) and acts as a SEB. Gate one is connected to a lumped-element resonator probed at a frequency $f_{\text{r}}$ for dispersive readout. The QDs under gates two and three are charge-sensed in the few-electron regime (red circles) and the gates are connected to fast pulse lines for spin manipulation. b) Resonator response at base temperature. c) Interdot charge transition (ICT) from (3,1) to (4,0) along which PSB is observed. The points \textbf{P} and \textbf{M} are used respectively to prepare and measure/initialise the spin state. d) Singlet (blue) and triplet (red) exemplary single shot traces with an energy diagram as insert showing the energy states close to the (3,1)-(4,0) anti-crossing.}
    \label{fig:PSB_Device}
\end{figure}

A favourable physical mechanism for single-shot readout is that of Pauli spin blockade (PSB) based on the large asymmetry in tunnel rates between two spin configurations.
In this second demonstration, the device consists of a silicon nanowire transistor with four wrap-around gates in series as shown in Fig.~\ref{fig:PSB_Device}a. The QD under gate 1 acts as a SEB with a larger, $\alpha=0.40$ due to the wrap around nature of the gates (See \S XIII for calculating $\alpha$, $T_e$ and $\gamma$). It is dispersively measured using an $LC$ lump-element resonator consisting of a 160~nH NbN superconducting spiral inductor and a capacitance $C_{\text{c}}$+$C_{\text{p}}$ of 250~fF (extracted from the natural frequency of the resonator, $f_\text{rf}=797$~MHz, in Fig.~\ref{fig:PSB_Device}b). Given the improved matching ($\beta=1.05$), internal quality factor ($Q_0=298$), resonator impedance ($Z_\text{r}=800$~$\Omega$), and higher frequency, this second implementation of the SEB reaches a $\tau_\text{m}$=170~ns (see Tab. I and \S XIV for the resonator analysis). We use the SEB to sense the charge state of a few-electron double quantum dot (DQD) under gates 2 and 3 using a PID feedback loop~\cite{yang2011dynamically}(See \S XV for its implementation). We observe a trapezoidal region in which PSB occurs both in the (1,1)-(2,0) and the (3,1)-(4,0) charge transitions from which we extract a 16~$\mu$eV and 195~$\mu$eV valley splittings, respectively (See \S XVI for valley splitting extraction). The larger (3,1)-(4,0) valley splitting aligns with previous observations in planar devices~\cite{West2019}. Due to the larger measurement window, we operate in the (3,1)-(4,0) charge transition region, as shown in Fig~\ref{fig:PSB_Device}c, acquired without feedback. The sensor operates in the strongly sensitive regime (See \S XVII). To determine the readout fidelity, we initialise by waiting at point $\textbf{M}$ for 5.1~ms, such that the system relaxes to the ground state, a singlet (4,0). To prepare a mixed singlet-triplet population, we pulse with a 100~ns ramp to point \textbf{P} and wait for 533~$\mu$s, with a 30~mT magnetic field applied. By pulsing back to point \textbf{M}, also with a 100~ns ramp, only the singlet is allowed to tunnel, which does so faster than the measurement bandwidth of 1.22~MHz, resulting in a sudden change in charge state. The triplet, being the excited state at point \textbf{M}, needs to relax to the singlet before it can tunnel, resulting in a delayed response of characteristic timescale $T_1$. We record the first 400~$\mu$s after the pulse to point \textbf{M} for 10,000 shots, for which we show exemplary traces for both singlet (blue) and triplets (red) in Fig~\ref{fig:PSB_Device}d. 

\begin{figure*}[ht]
    \centering
    \includegraphics[width=1\textwidth]{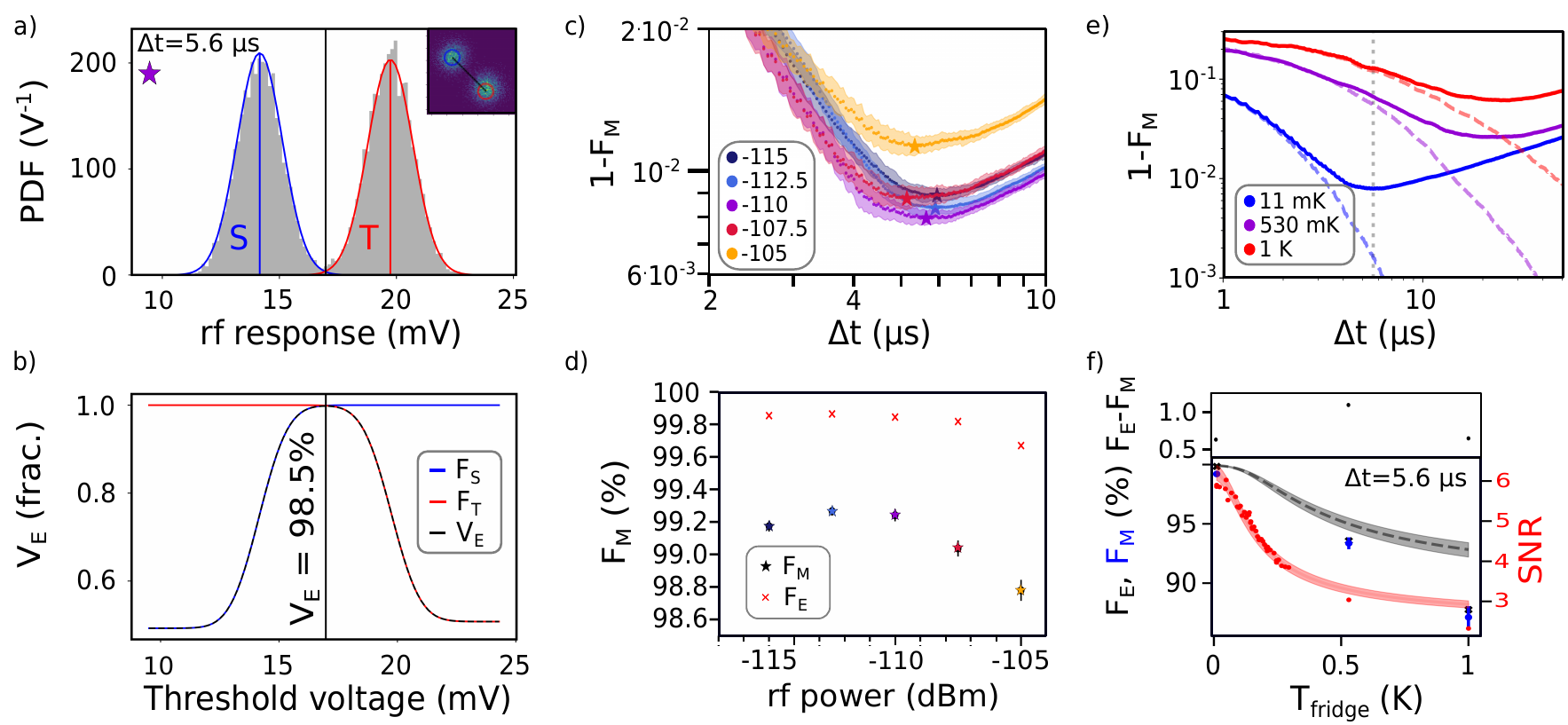}
    \caption{Spin readout fidelity. Data was acquired with a sampling rate of 20~MHz and a measurement bandwidth of 1.22~MHz. a) Histogram of 10,000 single shot measurements averaged over 5.6~$\mu$s with the fitted distribution of singlets (blue) and triplets (red), according to ref.~\cite{barthel2009rapid}. Insert shows the singlet triplet distribution in the IQ plane, from which the binomial distribution is obtained.  b) Readout visibility $V_{\text{E}}$ as a function of threshold voltage used to discern singlets from triplets. c) Measurement infidelity (1-$F_{\text{M}}$) with respect to measurement time $\Delta t$ for varying rf powers applied to the resonator. d) Readout fidelity as a function of rf power, reaching an optimum between -112.5 and -110~dBm. The red crosses correspond to the electrical fidelity. 
   e) Measurement infidelity (solid lines) and electrical infidelity (dashed lines) with respect to measurement time $\Delta t$ at varying temperatures measured at -110~dBm. The vertical dashed line is at $\Delta t=5.6$~$\mu$s. f) Measured readout fidelity (blue dots), electrical fidelity (black dots) and SEB signal to noise ratio (red dots) as a function of temperature for $\Delta t=5.6$~$\mu$s. The red line is a $(T_\text{e}^2+T_\text{fridge}^2)^{-1/2}$ fit to the SNR. The grey dashed line is the associated electrical fidelity, $F_\text{E}=[1+\text{erf}(\text{SNR}/(2\sqrt{2}))]/2$. Overhead insert highlighting $T_1$-induced errors.}
    \label{fig:PSB_readout}
\end{figure*}

By plotting the occurrence of average values of each trace for a given measurement time $\Delta t$, a bimodal probability distribution associated to the singlet and triplet outcomes appears, see Fig~\ref{fig:PSB_readout}a and \S XVIII. In this case, the data can be fit to a well-established model for PSB readout that includes Gaussian probabilities for the singlet and triplet outcomes as well as the contribution from triplet decays~\cite{barthel2009rapid}.
From the fitted parameters, we determine the optimum threshold voltage for a maximum visibility of 98.5\%, as shown in Fig.~\ref{fig:PSB_readout}b. This data analysis can be carried out as a function of measurement time $\Delta t$. In Fig.~\ref{fig:PSB_readout}c, the spin readout infidelity 1-F$_M$ decreases as $\Delta t$ increases due to a reduction of the measurement noise.
However, at longer time scales, triplets start decaying into the (4,0) charge configuration, which decreases the readout fidelity. We find $T_1=228.6\pm 0.5$~$\mu$s. Single-shot data sets were taken for varying rf powers, resulting in a maximum fidelity $F_\text{M}= 99.21\pm$0.03\% in 5.6~$\mu$s, as shown in Fig.~\ref{fig:PSB_readout}d. From the difference between the sensor-limited readout fidelity [considering $T_1\rightarrow\infty$ (red crosses)] and $F_\text{M}$ (stars), we determine the sources of error being 0.67$\pm$0.06\% due to $T_1$-induced errors and 0.23$\pm$0.09\% due to the sensor.

Finally, we explore the role of temperature. There has been recent interest in operating QDs at elevated temperatures~\cite{yang2020operation, petit2020universal} due to the additional cooling power that would facilitate the co-integration with classical electronic circuits and the development of an all-silicon quantum computing system~\cite{gonzalezzalba2020,Xue2021}.
We see in Fig.~\ref{fig:PSB_readout}e,f that $F_\text{M}$ decreases with increasing temperature to 97.39$\pm$0.18\% in 23.6~$\mu$s at 530~mK and to 93.90$\pm$0.28\% in 27.1~$\mu$s at 1~K. The prevailing factor in the decrease in fidelity is a reduction in the SNR of the SEB charge transition (red dots in Fig.~\ref{fig:PSB_readout}f) impacting the electrical infidelities (dotted lines in Fig.~\ref{fig:PSB_readout}e). This result also explains why the minimum infidelity occurs at higher $\Delta t$ with respect to base temperature. 
For $T_\text{fridge}\geq 500$~mK, the SNR decreases more rapidly than the expected $1/T_\text{fridge}$ dependence (red fit in panel f and see \S XIX) because the charge sensing shift of the SEB becomes smaller than its FWHM, leading to a fractional signal change, $\eta<1$. Summarising, sensor-induced errors are responsible for 12.2\% of the error at 1~K while $T_1$-induced errors for less than 1.0\%, see overhead insert in Fig.~\ref{fig:PSB_readout}f.

\section{Discussion} \label{sec: discussion}

We have determined that under asymmetric spin tunneling conditions, the two demonstrations would reach fault-tolerant readout fidelity in a few microseconds. These results highlight the benefits of PSB over tunneling to a reservoir but also opens the question of how best design a SEB to achieve fast readout. To answer this question, we calculate the technological parameters that determine $\tau_\mathrm{m}$. In the small signal regime (see \S I), the measurement rate (for a SNR=1) can be expressed as 

\begin{equation}
\label{eq:tau_m}
    \tau_\mathrm{m}^{-1}\propto\eta^2\frac{\beta}{(1+\beta)^2}\frac{(\alpha e)^2}{k_\text{B}T_\text{n}}Q_0Z_\text{r}\frac{f_\text{rf}^2} {\left(1+f_\text{rf}^2/\gamma^2 \right)^{2}}.
\end{equation}

We find that the charge sensing regime (quantified by $\eta$, the fractional change in $\Delta C$ due to a charge sensing event), the lever arm and the operation frequency have the highest impact with a quadratic dependence. However, $f_\text{rf}$ cannot be increased indefinitely otherwise electron tunneling to and from the SEB may not occur. Then low-loss, high-impedance resonators close to critical coupling are desired although $\tau_\mathrm{m}$ is first order insensitive to the coupling coefficient near $\beta=1$. Lastly, quantum-limited amplifiers are advantageous. In the context of our two demonstrations, see Table~\ref{table:1}, the lower noise temperature in the spin dependent tunneling experiment is compensated by the improved resonator specifications in conjunction with the higher $\alpha$ in the PSB-based demonstration. If a JPA were to be used in this instance, $\tau_\text{m}$ could be further reduced to $\approx 17$~ns. Considering this scenario, combined with a longer $T_1$ (1~ms) to reduce the impact of relaxation, a fidelity of 99.97\% in 1.2~$\mu$s could be achieved, based on the analysis in ref.~\cite{barthel2009rapid}. Further improvements can be achieved by using devices with smaller equivalent oxide thickness, and hence larger $\alpha$~\cite{Ahmed2018}, and microwave resonators~\cite{Zheng2019, Ibberson2021}. 

\begin{table}
\centering
\begin{tabular}{ |l|c|c|c|c| } 
 \hline
   \multirow{2}{5em}{\textbf{Type}} & \multirow{2}{5em}{\textbf{Parameter}} & \multicolumn{2}{|c|}{\textbf{Elzerman}} & \multirow{2}{4em}{\textbf{PSB}}  \\ \cline{3-4}
    & & \textbf{JPA off} & \textbf{JPA on} & \\  \hline\hline
    \multirow{5}{5em}{\textbf{Resonator}} & $\beta$ & 2.5 & 2.5 & 1.05 \\
    \cline{2-5}
    & $C_\text{tot}$(fF) & 461 & 461 & 250 \\ \cline{2-5}
    & $Q_0$ & 270
    & 270
    & 298 \\ \cline{2-5}
    & $Z_\text{r}(\Omega)$ & 519 & 519 & 800 \\ \cline{2-5}
    & $f_\text{r}$(MHz) & 665.5 & 668 & 797 \\ \hline \hline
    \multirow{4}{5em}{\textbf{SEB}} & $\alpha$ & 0.35 & 0.35 & 0.40 \\
    \cline{2-5}    
    & $\eta$ & 1 & 1 & 1 \\ \cline{2-5}
    & $\gamma$ (GHz) & 70
    & 74
    & $< 4.25$ \\ \cline{2-5}
    & $T_\text{e}$({mK}) & 137
    & 137
    & 115 \\  \hline \hline
    \textbf{Noise} & $T_\text{n}$(K) & 2.5 & 0.25(0.56*) & 2.5 \\ \hline\hline
    \textbf{Sensor} & $\tau_\mathrm{m}$($\mu$s) & 0.9 & 0.2 & 0.17 \\ \hline\hline
    \multirow{2}{5em}{\textbf{Spin mapping}} & $T_1$ (ms) & 5
    & 5
    & 0.23 \\ \cline{2-5}  & $t_\text{out}$ ($\mu$s) & 53 & 31 & $<$ 1** \\ \hline\hline
    \multirow{2}{6em}{\textbf{\small Benchmark}} &
    $F_\text{M}$(\%) & 99.1$^\dagger$ & 99.54$^{\dagger\dagger}$ & 99.21 \\ \cline{2-5}  & $\Delta t$ ($\mu$s) & 500$^\dagger$ & 250$^{\dagger\dagger}$ & 5.6 \\ \hline

\end{tabular}
\caption{Summary of parameters relevant for SEB charge sensing. *Indicates the noise temperature for the optimal SNR. **Refers to the singlet tunneling time in PSB. No Neural Networks applied $\dagger$ 98.85\% in 434~$\mu$s and $\dagger\dagger$ 99.45\% in 268~$\mu$s.}
\label{table:1}
\end{table}

\section{Conclusions and outlook} \label{sec: conclusions}

We have presented results at the state-of-the-art for spin readout using dispersive charge sensors and demonstrated high-fidelity readout in timescales much shorter than the coherence time of electron spins in silicon. The reduced footprint of the SEB compared to standard dissipative charge sensors, like the SET, will facilitate the development of highly-connected QD-based quantum processors, placing dispersive charge sensing at the forefront of readout methodologies for scalable spin-based quantum processors. 

\section{Methods}

\subsection{Fabrication details}\label{SubSec_Fab} 

Both devices are MOS transistors fabricated in an industrial cleanroom using 300~mm wafers of 7~nm silicon-on-insulator (SOI) with a $145$-nm-thick buried oxide. The gate oxide is 6~nm of thermal SiO$_2$ and the gate metal is 5~nm TiN and 50~nm poly-crystalline silicon. \textcolor{black}{Device 1} shown in Fig.~\ref{fig:Fig1}a has four gates on top of the nanowire, two on each side facing each other (See Fig.~\ref{fig:Fig1}a). The nanowire has a width of $W$=80~nm and a gate length of $L_\mathrm{G}=50$~nm. The separation between parallel gates is $S_\mathrm{H}=50$~nm, whereas between the gates facing each other is $S_\mathrm{V}=40$~nm. \textcolor{black}{Device 2} shown in Fig.~\ref{fig:PSB_Device}a has four wrap around-gates in series. The nanowire width is $W=80$~nm, the gate length $L_\mathrm{G}=50$~nm, and the gate separation is $S_\mathrm{H}=50$~nm. To form the gates, a hybrid Deep Ultra Violet/Electron Beam lithography and etching of the gate hard mask was performed before transferring the dense pattern into the rest of the stack. After gate-etching, the nanowire is covered by 34 nm-wide $\mathrm{Si_3N_4}$ spacers. On one hand, the spacer separates the reservoirs from the central part of the intrinsic nanowire,  protecting the intrinsic silicon from the posterior ion implantation which defines the reservoirs.  And, on the other hand, it also covers the split between the independent gates since the spacer length is larger than half of the inter-gate gap.  The reservoirs are then n-doped by Arsenic/Phosphorus implantation. The process is completed after an activation spike anneal, salicidation (NiPtSi), contacts and metallization.

\textcolor{black}{In device 1}, up to four QDs can be formed on the upper corners of the nanowire, two at each side of the nanowire forming a $2\times 2$ configuration. {In device 2}, three QDs can be formed in a $1\times 3$ configuration. Gate 4 was not working in this particular demonstration. The dots' electrochemical potential can be controlled by the voltage applied to the gates above them. 
For further control, the silicon substrate can be used as a back-gate and an overarching metal line as a top gate (only in device 1).

\subsection{Measurement setup}\label{SubSec_Setup} 
Measurements were performed at base temperature of a dilution refrigerator (setup 1(2) at $15(11)$~mK). DC voltages to the gates and electron reservoirs were delivered through filtered cryogenic loom. In setup 1, the inter-dot and dot-to-reservoir tunneling rates can be modified by the DC voltage applied to a top metallic gate~\cite{Ansaloni2020}. The experiments were performed at a constant voltage of $V_{\mathrm{top}}=7.5V$ applied to the metal line. 

The devices electrodes are connected to the PCB contacts via on-chip aluminium bond wires. On-PCB bias tees are used to combine the DC signals with the radio-frequency signal for gate-based readout and the fast pulses, which are delivered through attenuated and filtered coaxial lines. The bias tee acts on the pulses sent as a high pass filter. This effect was compensated by pulse engineering using the inverse of the filter transfer function, such that after passing through the bias tee, the pulses had the desired lineshape. 

The resonator for reflectometry is formed in setup 1(2) by an $124(160)$~nH NbN planar spiral inductor~\cite{Ahmed2018} placed in parallel to the parasitic capacitance to ground of the PCB and the device. It is capacitvely coupled to a 50~$\Omega$ transmission line via a coupling capacitor 50(57)~fF. The gate connected to the resonator is biased through a low pass filter on the PCB formed by a 100~k$\Omega$ resistor in series with the inductor and a 100~pF capacitor to ground. The PCB is made from $0.8$~mm thick RO4003C with immersion silver finish. On its way out of the fridge, the reflected rf-signal is first amplified by 26~dB at $4$~K (LNF-LNC0.6$\_$2A) and further amplified at room temperature. Then, the reflected signal magnitude and phase are obtained using quadrature demodulation (Polyphase AD0540b) and measured using a digitiser in setup 1 (Spectrum M4i.4451-x4) and an oscilloscope in setup 2 (HDO4054A Lecroy). The measurements taken using a JPA have an additional amplification of 16.7~dB at 105~mK~\cite{Simbierowicz2018}.

\section{Data Availability}

The data that support the plots within this paper and other findings of this study are available from the corresponding authors upon reasonable request.

\section{Acknowledgements}

This research was supported by European Union’s Horizon 2020 research and innovation programme under grant agreement no.\ 951852 (QLSI), and by the UK's Engineering and Physical Sciences Research Council (EPSRC) via the Cambridge NanoDTC (EP/L015978/1), CDT in Delivering Quantum Technologies (EP/L015242/1), QUES2T (EP/N015118/1), and the Hub in Quantum Computing and Simulation (EP/T001062/1).
V.N.C.T.\ is a Telefónica British-Spanish society scholar. 
M.F.G.-Z.\ is a UKRI Future Leaders Fellow (MR/V023284/1). 

\section{Author contributions}

V.C-T performed the experiments on spin-dependent tunneling with assistance of S.S. and M.F.G.Z. G.A.O. performed the experiments on Pauli spin-blockade with assistance of T.L, F.R.M and M.F.G.Z.. S.S., G.A.O and M.F.G.Z. designed the PCBs. L.H., B.B. and M.V. designed and fabricated the devices. C.L. developed the code to analyse the spin-dependent tunneling data. D.W. performed the Neural Network analysis of the spin-dependent tunneling shots. A.H and I.S. designed and fabricated the JPA. N.S. and J.A.W.R. provided the NbN films. D.J.I and L.I. manufactured the spiral NbN inductors. M.F.G.Z developed the theory of SEB readout. V.C-T., G.A.O and M.F.G-Z wrote the manuscript with input from all co-authors. M.F.G.Z. and J.J.L.M supervised the spin-dependent tunneling experiment. A.L., C.G.S and M.F.G.Z supervised the Pauli spin blockade experiment. 

\section{Competing interests}

G.A.O, V.N.C-T, D.W., M.A.F, C.L, S.S., J.J.L.M and M.F.G.-Z. are supported by Quantum Motion, a start-up developing silicon-based quantum computing. All other authors declare no competing interests.

\section {Additional information}



Correspondence and request for materials should be addressed to J.J.L.M (john@quantummotion.tech) and M.F.G.Z (fernando@quantummotion.tech).


\begin{thebibliography}{10}
\expandafter\ifx\csname url\endcsname\relax
  \def\url#1{\texttt{#1}}\fi
\expandafter\ifx\csname urlprefix\endcsname\relax\def\urlprefix{URL }\fi
\providecommand{\bibinfo}[2]{#2}
\providecommand{\eprint}[2][]{\url{#2}}

\bibitem{xue2021a}
\bibinfo{author}{Xue, X.} \emph{et~al.}
\newblock \bibinfo{title}{Quantum logic with spin qubits crossing the surface
  code threshold}.
\newblock \emph{\bibinfo{journal}{Nature}} \textbf{\bibinfo{volume}{601}},
  \bibinfo{pages}{343--347} (\bibinfo{year}{2022}).

\bibitem{Noiri2021}
\bibinfo{author}{Noiri, A.} \emph{et~al.}
\newblock \bibinfo{title}{Fast universal quantum gate above the fault-tolerance
  threshold in silicon}.
\newblock \emph{\bibinfo{journal}{Nature}} \textbf{\bibinfo{volume}{601}},
  \bibinfo{pages}{338--342} (\bibinfo{year}{2022}).

\bibitem{Mills2021}
\bibinfo{author}{Mills, A.~R.} \emph{et~al.}
\newblock \bibinfo{title}{Two-qubit silicon quantum processor with operation
  fidelity exceeding 99\%}.
\newblock \emph{\bibinfo{journal}{Preprint at
  \tt{https://arxiv.org/abs/2111.11937}}}  (\bibinfo{year}{2021}).

\bibitem{Veldhorst2017}
\bibinfo{author}{Veldhorst, M.}, \bibinfo{author}{Eenink, H. G.~J.},
  \bibinfo{author}{Yang, C.~H.} \& \bibinfo{author}{Dzurak, A.~S.}
\newblock \bibinfo{title}{{Silicon CMOS architecture for a spin-based quantum
  computer}}.
\newblock \emph{\bibinfo{journal}{Nat. Commun.}} \textbf{\bibinfo{volume}{8}},
  \bibinfo{pages}{1766} (\bibinfo{year}{2017}).

\bibitem{Boter2021}
\bibinfo{author}{Boter, J.~M.} \emph{et~al.}
\newblock \bibinfo{title}{{The spider-web array--a sparse spin qubit array}}.
\newblock \emph{\bibinfo{journal}{Preprint at
  \tt{https://arxiv.org/abs/2110.00189}}}  (\bibinfo{year}{2021}).

\bibitem{Maurand2016}
\bibinfo{author}{Maurand, R.} \emph{et~al.}
\newblock \bibinfo{title}{{A CMOS silicon spin qubit}}.
\newblock \emph{\bibinfo{journal}{Nat. Commun.}} \textbf{\bibinfo{volume}{7}},
  \bibinfo{pages}{13575} (\bibinfo{year}{2016}).

\bibitem{Zwerver2021}
\bibinfo{author}{Zwerver, A. M.~J.} \emph{et~al.}
\newblock \bibinfo{title}{Qubits made by advanced semiconductor manufacturing}.
\newblock \emph{\bibinfo{journal}{Preprint at
  \tt{https://arxiv.org/abs/2101.12650}}}  (\bibinfo{year}{2021}).

\bibitem{Ruffino2021a}
\bibinfo{author}{Ruffino, A.} \emph{et~al.}
\newblock \bibinfo{title}{A cryo-cmos chip that integrates silicon quantum dots
  and multiplexed dispersive readout electronics}.
\newblock \emph{\bibinfo{journal}{Nat. Electron.}}
  \textbf{\bibinfo{volume}{5}}, \bibinfo{pages}{53--59} (\bibinfo{year}{2022}).

\bibitem{McArdle2019}
\bibinfo{author}{McArdle, S.}, \bibinfo{author}{Yuan, X.} \&
  \bibinfo{author}{Benjamin, S.}
\newblock \bibinfo{title}{{Error-Mitigated Digital Quantum Simulation}}.
\newblock \emph{\bibinfo{journal}{Phys. Rev. Lett.}}
  \textbf{\bibinfo{volume}{122}}, \bibinfo{pages}{180501}
  (\bibinfo{year}{2019}).

\bibitem{Botelho2021}
\bibinfo{author}{Botelho, L.} \emph{et~al.}
\newblock \bibinfo{title}{{Error mitigation for variational quantum algorithms
  through mid-circuit measurements}}.
\newblock \emph{\bibinfo{journal}{arXiv Prepr. arXiv2108.10927}}
  (\bibinfo{year}{2021}).
\newblock \eprint{2108.10927}.

\bibitem{Wan2019}
\bibinfo{author}{Wan, Y.} \emph{et~al.}
\newblock \bibinfo{title}{{Quantum gate teleportation between separated qubits
  in a trapped-ion processor}}.
\newblock \emph{\bibinfo{journal}{Science (80-. ).}}
  \textbf{\bibinfo{volume}{364}}, \bibinfo{pages}{875--878}
  (\bibinfo{year}{2019}).

\bibitem{Curry2019}
\bibinfo{author}{Curry, M.~J.} \emph{et~al.}
\newblock \bibinfo{title}{Single-shot readout performance of two
  heterojunction-bipolar-transistor amplification circuits at millikelvin
  temperatures}.
\newblock \emph{\bibinfo{journal}{Sci. Rep.}} \textbf{\bibinfo{volume}{9}},
  \bibinfo{pages}{16976} (\bibinfo{year}{2019}).

\bibitem{Connors2020}
\bibinfo{author}{Connors, E.~J.}, \bibinfo{author}{Nelson, J.} \&
  \bibinfo{author}{Nichol, J.~M.}
\newblock \bibinfo{title}{Rapid high-fidelity spin-state readout in si / si -
  ge quantum dots via rf reflectometry}.
\newblock \emph{\bibinfo{journal}{Phys. Rev. App.}}
  \textbf{\bibinfo{volume}{13}}, \bibinfo{pages}{024019}
  (\bibinfo{year}{2020}).

\bibitem{Petersson2010}
\bibinfo{author}{Petersson, K.~D.} \emph{et~al.}
\newblock \bibinfo{title}{Charge and spin state readout of a double quantum dot
  coupled to a resonator}.
\newblock \emph{\bibinfo{journal}{Nano Lett.}} \textbf{\bibinfo{volume}{10}},
  \bibinfo{pages}{2789--2793} (\bibinfo{year}{2010}).

\bibitem{Colless2013}
\bibinfo{author}{Colless, J.~I.} \emph{et~al.}
\newblock \bibinfo{title}{Dispersive readout of a few-electron double quantum
  dot with fast rf gate sensors}.
\newblock \emph{\bibinfo{journal}{Phys. Rev. Lett.}}
  \textbf{\bibinfo{volume}{110}}, \bibinfo{pages}{46805}
  (\bibinfo{year}{2013}).

\bibitem{GonzalezZalba2015}
\bibinfo{author}{Gonzalez-Zalba, M.~F.}, \bibinfo{author}{Barraud, S.},
  \bibinfo{author}{Ferguson, A.~J.} \& \bibinfo{author}{Betz, A.~C.}
\newblock \bibinfo{title}{{Probing the limits of gate-based charge sensing}}.
\newblock \emph{\bibinfo{journal}{Nat. Commun.}} \textbf{\bibinfo{volume}{6}},
  \bibinfo{pages}{6084} (\bibinfo{year}{2015}).

\bibitem{House2016}
\bibinfo{author}{House, M.~G.} \emph{et~al.}
\newblock \bibinfo{title}{High-sensitivity charge detection with a single-lead
  quantum dot for scalable quantum computation}.
\newblock \emph{\bibinfo{journal}{Phys. Rev. App.}}
  \textbf{\bibinfo{volume}{6}}, \bibinfo{pages}{044016} (\bibinfo{year}{2016}).

\bibitem{Urdampilleta2019}
\bibinfo{author}{Urdampilleta, M.} \emph{et~al.}
\newblock \bibinfo{title}{{Gate-based high fidelity spin readout in a CMOS
  device}}.
\newblock \emph{\bibinfo{journal}{Nat. Nanotechnol.}}
  \textbf{\bibinfo{volume}{14}}, \bibinfo{pages}{737--741}
  (\bibinfo{year}{2019}).

\bibitem{Chanrion2020}
\bibinfo{author}{Chanrion, E.} \emph{et~al.}
\newblock \bibinfo{title}{{Charge detection in an array of CMOS quantum dots}}.
\newblock \emph{\bibinfo{journal}{Phys. Rev. App.}}
  \textbf{\bibinfo{volume}{14}}, \bibinfo{pages}{024066}
  (\bibinfo{year}{2020}).

\bibitem{Ansaloni2020}
\bibinfo{author}{Ansaloni, F.} \emph{et~al.}
\newblock \bibinfo{title}{Single-electron operations in a foundry-fabricated
  array of quantum dots}.
\newblock \emph{\bibinfo{journal}{Nat. Commun.}} \textbf{\bibinfo{volume}{11}},
  \bibinfo{pages}{6399} (\bibinfo{year}{2020}).

\bibitem{CirianoTejel2021}
\bibinfo{author}{Ciriano-Tejel, V.~N.} \emph{et~al.}
\newblock \bibinfo{title}{{Spin Readout of a CMOS Quantum Dot by Gate
  Reflectometry and Spin-Dependent Tunneling}}.
\newblock \emph{\bibinfo{journal}{PRX Quantum}} \textbf{\bibinfo{volume}{2}},
  \bibinfo{pages}{010353} (\bibinfo{year}{2021}).

\bibitem{Borjans2021}
\bibinfo{author}{Borjans, F.}, \bibinfo{author}{Mi, X.} \&
  \bibinfo{author}{Petta, J.}
\newblock \bibinfo{title}{Spin digitizer for high-fidelity readout of a
  cavity-coupled silicon triple quantum dot}.
\newblock \emph{\bibinfo{journal}{Phys. Rev. App.}}
  \textbf{\bibinfo{volume}{15}}, \bibinfo{pages}{044052}
  (\bibinfo{year}{2021}).

\bibitem{Vijay2009}
\bibinfo{author}{Vijay, R.}, \bibinfo{author}{Devoret, M.~H.} \&
  \bibinfo{author}{Siddiqi, I.}
\newblock \bibinfo{title}{{Invited Review Article: The Josephson bifurcation
  amplifier}}.
\newblock \emph{\bibinfo{journal}{Rev. Sci. Instrum.}}
  \textbf{\bibinfo{volume}{80}}, \bibinfo{pages}{111101}
  (\bibinfo{year}{2009}).
  
\bibitem{Vigneau2022}
\bibinfo{author}{Vigneau, F.}  \emph{et~al.}
\newblock \bibinfo{title}{{Probing quantum devices with radio-frequency reflectometry}}.
\newblock \emph{\bibinfo{journal}{Preprint at
  \tt{https://arxiv.org/abs/2202.10516}}}  (\bibinfo{year}{2022}).

\bibitem{Keith2019}
\bibinfo{author}{Keith, D.} \emph{et~al.}
\newblock \bibinfo{title}{Single-shot spin readout in semiconductors near the
  shot-noise sensitivity limit}.
\newblock \emph{\bibinfo{journal}{Phys. Rev. X}} \textbf{\bibinfo{volume}{9}},
  \bibinfo{pages}{41003} (\bibinfo{year}{2019}).

\bibitem{Schaal2020}
\bibinfo{author}{Schaal, S.} \emph{et~al.}
\newblock \bibinfo{title}{Fast gate-based readout of silicon quantum dots using
  josephson parametric amplification}.
\newblock \emph{\bibinfo{journal}{Phys. Rev. Lett.}}
  \textbf{\bibinfo{volume}{124}}, \bibinfo{pages}{67701}
  (\bibinfo{year}{2020}).

\bibitem{Stehlik2015}
\bibinfo{author}{Stehlik, J.} \emph{et~al.}
\newblock \bibinfo{title}{Fast charge sensing of a cavity-coupled double
  quantum dot using a josephson parametric amplifier}.
\newblock \emph{\bibinfo{journal}{Phys. Rev. App.}}  (\bibinfo{year}{2015}).

\bibitem{Elzerman2004}
\bibinfo{author}{Elzerman, J.~M.} \emph{et~al.}
\newblock \bibinfo{title}{Single-shot read-out of an individual electron spin
  in a quantum dot}.
\newblock \emph{\bibinfo{journal}{Nature}} \textbf{\bibinfo{volume}{430}}
  (\bibinfo{year}{2004}).

\bibitem{Barthel2010}
\bibinfo{author}{Barthel, C.} \emph{et~al.}
\newblock \bibinfo{title}{Fast sensing of double-dot charge arrangement and
  spin state with a radio-frequency sensor quantum dot}.
\newblock \emph{\bibinfo{journal}{Phys. Rev. B}} \textbf{\bibinfo{volume}{81}},
  \bibinfo{pages}{161308(R)} (\bibinfo{year}{2010}).

\bibitem{Struck2021}
\bibinfo{author}{Struck, T.} \emph{et~al.}
\newblock \bibinfo{title}{{Robust and fast post-processing of single-shot spin
  qubit detection events with a neural network}}.
\newblock \emph{\bibinfo{journal}{Sci. Rep.}} \textbf{\bibinfo{volume}{11}},
  \bibinfo{pages}{16203} (\bibinfo{year}{2021}).

\bibitem{Matsumoto2020}
\bibinfo{author}{Matsumoto, Y.} \emph{et~al.}
\newblock \bibinfo{title}{{Noise-robust classification of single-shot electron
  spin readouts using a deep neural network}}.
\newblock \emph{\bibinfo{journal}{npj Quantum Inf.}}
  \textbf{\bibinfo{volume}{7}}, \bibinfo{pages}{136} (\bibinfo{year}{2021}).

\bibitem{yang2011dynamically}
\bibinfo{author}{Yang, C.}, \bibinfo{author}{Lim, W.},
  \bibinfo{author}{Zwanenburg, F.} \& \bibinfo{author}{Dzurak, A.}
\newblock \bibinfo{title}{Dynamically controlled charge sensing of a
  few-electron silicon quantum dot}.
\newblock \emph{\bibinfo{journal}{AIP Advances}} \textbf{\bibinfo{volume}{1}},
  \bibinfo{pages}{042111} (\bibinfo{year}{2011}).

\bibitem{West2019}
\bibinfo{author}{West, A.} \emph{et~al.}
\newblock \bibinfo{title}{Gate-based single-shot readout of spins in silicon}.
\newblock \emph{\bibinfo{journal}{Nat. Nanotechnol.}}
  \textbf{\bibinfo{volume}{14}}, \bibinfo{pages}{437--441}
  (\bibinfo{year}{2019}).

\bibitem{barthel2009rapid}
\bibinfo{author}{Barthel, C.}, \bibinfo{author}{Reilly, D.},
  \bibinfo{author}{Marcus, C.~M.}, \bibinfo{author}{Hanson, M.} \&
  \bibinfo{author}{Gossard, A.}
\newblock \bibinfo{title}{Rapid single-shot measurement of a singlet-triplet
  qubit}.
\newblock \emph{\bibinfo{journal}{Phys. Rev. Lett.}}
  \textbf{\bibinfo{volume}{103}}, \bibinfo{pages}{160503}
  (\bibinfo{year}{2009}).

\bibitem{yang2020operation}
\bibinfo{author}{Yang, C.~H.} \emph{et~al.}
\newblock \bibinfo{title}{Operation of a silicon quantum processor unit cell
  above one kelvin}.
\newblock \emph{\bibinfo{journal}{Nature}} \textbf{\bibinfo{volume}{580}},
  \bibinfo{pages}{350--354} (\bibinfo{year}{2020}).

\bibitem{petit2020universal}
\bibinfo{author}{Petit, L.} \emph{et~al.}
\newblock \bibinfo{title}{Universal quantum logic in hot silicon qubits}.
\newblock \emph{\bibinfo{journal}{Nature}} \textbf{\bibinfo{volume}{580}},
  \bibinfo{pages}{355--359} (\bibinfo{year}{2020}).

\bibitem{gonzalezzalba2020}
\bibinfo{author}{Gonzalez-Zalba, M.~F.} \emph{et~al.}
\newblock \bibinfo{title}{Scaling silicon-based quantum computing using cmos
  technology}.
\newblock \emph{\bibinfo{journal}{Nat. Electron.}}
  \textbf{\bibinfo{volume}{4}}, \bibinfo{pages}{872--884}
  (\bibinfo{year}{2021}).

\bibitem{Xue2021}
\bibinfo{author}{Xue, X.} \emph{et~al.}
\newblock \bibinfo{title}{{CMOS-based cryogenic control of silicon quantum
  circuits}}.
\newblock \emph{\bibinfo{journal}{Nature}} \textbf{\bibinfo{volume}{593}},
  \bibinfo{pages}{205--210} (\bibinfo{year}{2021}).

\bibitem{Ahmed2018}
\bibinfo{author}{Ahmed, I.} \emph{et~al.}
\newblock \bibinfo{title}{Radio-frequency capacitive gate-based sensing}.
\newblock \emph{\bibinfo{journal}{Phys. Rev. App.}}
  \textbf{\bibinfo{volume}{10}}, \bibinfo{pages}{014018}
  (\bibinfo{year}{2018}).

\bibitem{Zheng2019}
\bibinfo{author}{Zheng, G.} \emph{et~al.}
\newblock \bibinfo{title}{Rapid gate-based spin read-out in silicon using an
  on-chip resonator}.
\newblock \emph{\bibinfo{journal}{Nat. Nanotechnol.}}
  \textbf{\bibinfo{volume}{14}}, \bibinfo{pages}{742--746}
  (\bibinfo{year}{2019}).

\bibitem{Ibberson2021}
\bibinfo{author}{Ibberson, D.~J.} \emph{et~al.}
\newblock \bibinfo{title}{Large dispersive interaction between a {CMOS} double
  quantum dot and microwave photons}.
\newblock \emph{\bibinfo{journal}{PRX Quantum}} \textbf{\bibinfo{volume}{2}},
  \bibinfo{pages}{020315} (\bibinfo{year}{2021}).

\bibitem{Simbierowicz2018}
\bibinfo{author}{Simbierowicz, S.} \emph{et~al.}
\newblock \bibinfo{title}{A flux-driven josephson parametric amplifier for
  sub-Ghz frequencies fabricated with side-wall passivated spacer junction
  technology}.
\newblock \emph{\bibinfo{journal}{Supercond. Sci. Technol.}}
  \textbf{\bibinfo{volume}{31}} (\bibinfo{year}{2018}).

\end{thebibliography}

\end{document}


\title{Fast high-fidelity single-shot readout of spins in silicon using a single-electron box}
	
\author{G. A. Oakes}
 \email{These authors contributed equally to this work}
 \affiliation{Cavendish Laboratory, University of Cambridge, J.J. Thomson Avenue, Cambridge CB3 0HE, United Kingdom}
 \affiliation{Quantum Motion, 9 Sterling Way, London N7 9HJ, United Kingdom}
\author{V.N. Ciriano-Tejel}%
 \email{These authors contributed equally to this work}
 \affiliation{Quantum Motion, 9 Sterling Way, London N7 9HJ, United Kingdom}
 \affiliation{London Centre for Nanotechnology, University College London, London WC1H 0AH, United Kingdom}
\author{D. Wise}%
 \affiliation{Quantum Motion, 9 Sterling Way, London N7 9HJ, United Kingdom}\affiliation{London Centre for Nanotechnology, University College London, London WC1H 0AH, United Kingdom}
\author{M. A. Fogarty}%
 \affiliation{Quantum Motion, 9 Sterling Way, London N7 9HJ, United Kingdom}
 \affiliation{London Centre for Nanotechnology, University College London, London WC1H 0AH, United Kingdom}
 \author{T. Lundberg}%
 \affiliation{Cavendish Laboratory, University of Cambridge, J.J. Thomson Avenue, Cambridge CB3 0HE, United Kingdom}
 \affiliation{Hitachi Cambridge Laboratory, J.J. Thomson Avenue, Cambridge CB3 0HE, United Kingdom}
\author{C. Lain\'{e}}%
 \affiliation{Quantum Motion, 9 Sterling Way, London N7 9HJ, United Kingdom}
\affiliation{London Centre for Nanotechnology, University College London, London WC1H 0AH, United Kingdom}
 \author{S. Schaal}%
 \affiliation{Quantum Motion, 9 Sterling Way, London N7 9HJ, United Kingdom}
\affiliation{London Centre for Nanotechnology, University College London, London WC1H 0AH, United Kingdom}
 \author{F.~Martins}%
 \affiliation{Hitachi Cambridge Laboratory, J.J. Thomson Avenue, Cambridge CB3 0HE, United Kingdom}
\author{D. J. Ibberson}%
 \affiliation{Quantum Engineering Technology Labs, University of Bristol, Tyndall Avenue, Bristol BS8 1FD, United Kingdom}
 \affiliation{Hitachi Cambridge Laboratory, J.J. Thomson Avenue, Cambridge CB3 0HE, United Kingdom}
\author{L. Hutin}%
 \affiliation{CEA, LETI, Minatec Campus, F-38054 Grenoble, France}
\author{B. Bertrand}%
 \affiliation{CEA, LETI, Minatec Campus, F-38054 Grenoble, France}
\author{N. Stelmashenko}%
 \affiliation{Department of Materials Science and Metallurgy, University of Cambridge,
27 Charles Babbage Road, Cambridge CB3 0FS, United Kingdom}
\author{J. A. W. Robinson}%
 \affiliation{Department of Materials Science and Metallurgy, University of Cambridge,
27 Charles Babbage Road, Cambridge CB3 0FS, United Kingdom}
\author{L. Ibberson}%
 \affiliation{Hitachi Cambridge Laboratory, J.J. Thomson Avenue, Cambridge CB3 0HE, United Kingdom}
\author{A.~Hashim}%
 \affiliation{Quantum Nanoelectronics Laboratory, Dept. of Physics, Univ. of California, Berkeley CA 94720, USA}
\author{I. Siddiqi}%
 \affiliation{Quantum Nanoelectronics Laboratory, Dept. of Physics, Univ. of California, Berkeley CA 94720, USA}
\author{A. Lee}%
 \affiliation{Cavendish Laboratory, University of Cambridge, J.J. Thomson Avenue, Cambridge CB3 0HE, United Kingdom}
\author{M. Vinet}%
 \affiliation{CEA, LETI, Minatec Campus, F-38054 Grenoble, France}
\author{C. G. Smith}
 \affiliation{Cavendish Laboratory, University of Cambridge, J.J. Thomson Avenue, Cambridge CB3 0HE, United Kingdom}
 \affiliation{Hitachi Cambridge Laboratory, J.J. Thomson Avenue, Cambridge CB3 0HE, United Kingdom}
\author{J.J.L. Morton}
 \email{john@quantummotion.tech}
 \affiliation{Quantum Motion, 9 Sterling Way, London N7 9HJ, United Kingdom}
\affiliation{London Centre for Nanotechnology, University College London, London WC1H 0AH, United Kingdom}
\author{M. F. Gonzalez-Zalba}
 \email{fernando@quantummotion.tech}
 \affiliation{Quantum Motion, 9 Sterling Way, London N7 9HJ, United Kingdom}

	\date{\today}

\maketitle

\section{Single-Electron Box optimisation}
\label{sup:subsecgamma}
In this section, we derive the Eq.~(3) in the main text. The SNR of the method is defined as

\begin{equation}\label{eq:SNR}
    \text{SNR}=\frac{\Delta P_\text{rf}}{P_\text{n}}=|\Delta\Gamma|^2\frac{P_\text{in}}{P_\text{n}}
\end{equation}

\noindent where $\Delta\Gamma$ is the change in reflection coefficient between the two spin states and $P_\text{in(n)}$ is the input(noise) power. In the small signal regime, where the product of the loaded quality factor and fractional change in capacitance is $Q_\text{L}\Delta C_\text{D}/C_\text{tot}\ll 1$, $\Delta\Gamma$ can be calculated as

\begin{equation}\label{eq:deltaGamma}
    \Delta\Gamma= \Delta C_\text{D}\times\left.\frac{\partial\Gamma}{\partial C_\text{D}}\right|_{f_\text{rf}=f_\text{res}}= i\frac{2\beta}{(1+\beta)^2}Q_0\frac{\Delta C_\mathrm{D}}{C_\mathrm{tot}},
\end{equation}

\noindent where $f_\text{res}$ is the natural frequency of oscillation, $\beta$ is the coupling coefficient and $Q_0$ is the internal Q-factor~\cite{Gonzalez-Zalba2018}.
The change in capacitance due to a charge sensing event in the low-power and thermally-broadened regime is~\cite{Gonzalez-Zalba2015} 

\begin{equation}\label{eq:deltaC}
    \Delta C_\mathrm{D}=\eta\frac{(\alpha e)^2}{2k_\text{B}T_\text{e}}\frac{1}{1+\left( f_\text{rf}/\gamma\right)^2}. 
\end{equation}

\noindent where $\eta$ is the fractional change in capacitance due to a charge sensing event (bounded between 0 and 1), $\alpha$ is the gate lever arm, $e$ the charge quantum, $k_\text{B}$ the Boltzmann constant, $T_\text{e}$ is the electron temperature and $\gamma$ is the SEB-reservoir tunnel rate. However, when driven at higher powers, the maximum value of Eq.~\eqref{eq:deltaC} can be reduced due to power broadening effects. In this limit, the capacitance can be calculated following the approach in ref.~\cite{Maman2020} that considers the adibatic limit (where rf-induced excitation and inelastic relaxation processes can be neglected). We find the change in capacitance

\begin{equation}\label{eq:deltaC2}
    \Delta C_\mathrm{D}=\eta\frac{2\alpha e}{\pi V_\text{dev}}\frac{1}{1+\left( f_\text{rf}/\gamma\right)^2}f_\text{c}(x), 
\end{equation}

\noindent where $V_\text{dev}$ is the voltage amplitude of the oscillatory voltage arriving at the gate of the SEB, $x=\frac{\alpha eV_\text{dev}}{k_\text{B}T}$, and $f_\text{c}$ is a dimensionless function of the form

\begin{equation}\label{eq:deltaC3}
    f_\text{c}(x)=\frac{1}{2}\int^{1/f_\text{rf}}_0\frac{\sin(2\pi f_\text{rf} t) dt}{1+\exp\left[-x\sin(2\pi f_\text{rf} t)\right]}. 
\end{equation}

\noindent which increases monotonically as a function of $V_\text{dev}$ until it saturates to the value of 1. Next, we calculate the relationship between $P_\text{in}$ and $V_\text{dev}$. For the capacitively coupled parallel $LCR$ resonator used in the main text, we find that

\begin{equation}\label{eq:power}
    P_\mathrm{in}=\frac{V_\text{dev}^2}{R}\frac{(1+\beta)^2}{4\beta}.
\end{equation}

Here $R$ corresponds to the resonator losses, a resistor in parallel with the inductor, parasitic capacitance and SEB. We then substitute equations Eq.~\eqref{eq:deltaGamma},\eqref{eq:deltaC2} and \eqref{eq:power} in Eq.~\eqref{eq:SNR} and find Eq.~(3) in the main text:

\begin{equation}
    \tau_\mathrm{m}^{-1}=32\eta^2\frac{\beta}{(1+\beta)^2}\frac{(\alpha e)^2}{k_\text{B}T_\text{n}}Q_0Z_\text{r}f_\text{rf}^2 \left[1+f_\text{rf}^2/\gamma^2 \right]^{-2}.
\end{equation}

Here we have used $Q_0=R\sqrt{C_\text{tot}/L}$ and defined the loaded resonator impedance $Z_\text{r}=\sqrt{L/C_\text{tot}}$. Further, we have used $P_\mathrm{n}=k_\text{B}T_\text{n}/(2\tau$), with $\tau$ being the integration time and $\tau_\mathrm{m}$ the integration time for SNR=1. We note the noise temperature on the JPA is gain dependent. If the reflected power from the resonator is larger than the 1~dB compression point, the gain drops with a consequent increase in noise temperature.






\section{Resonator parameter extraction}
\label{SupSec_optimisation}
\begin{figure}
    \centering
    \includegraphics[width=0.9\linewidth]{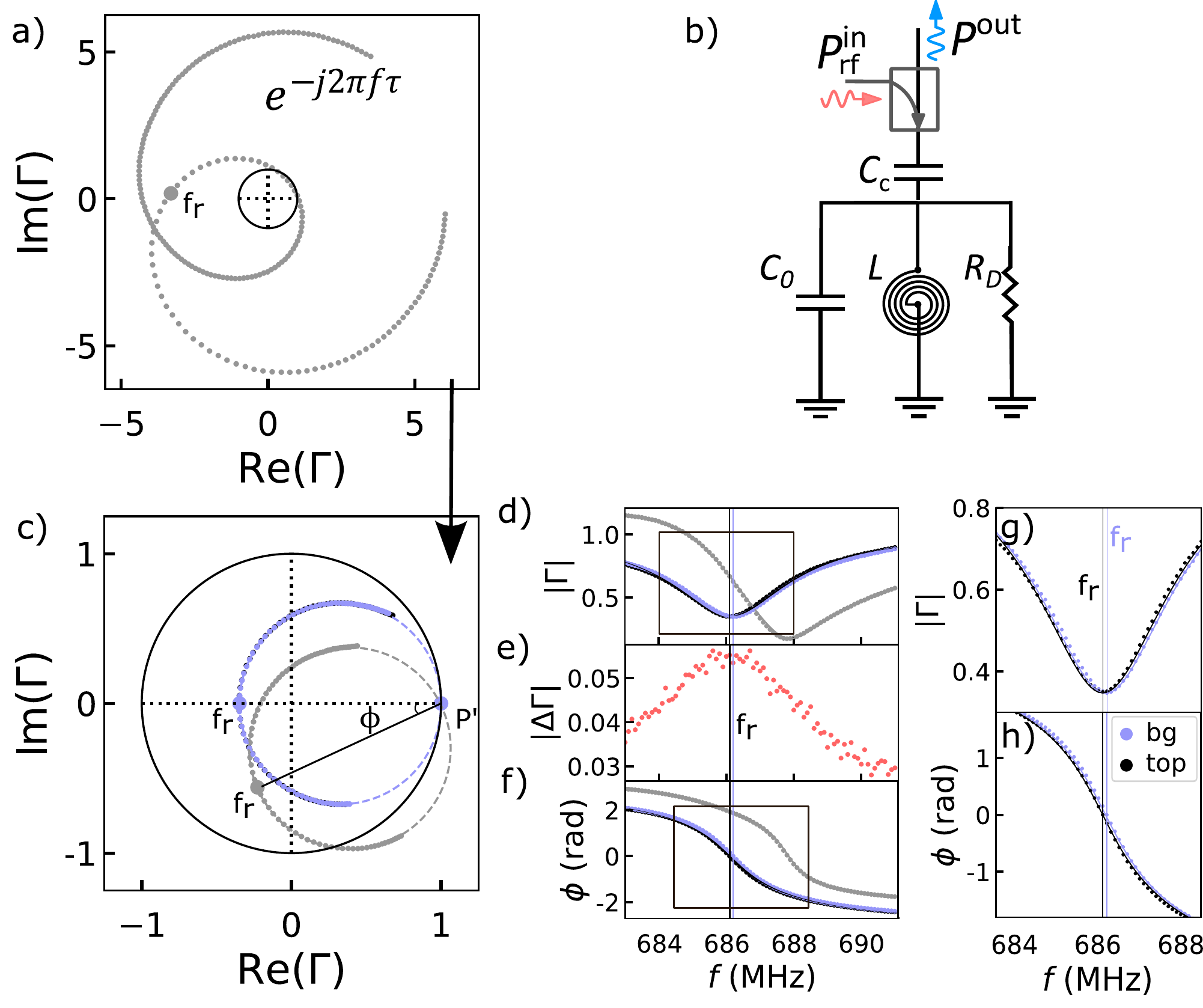}
    \caption{a) $\Gamma$ raw data acquired from a $\sqrt{S_{21}}$ measurement between the ports $P\mathrm{_{rf}^{in}}$ and $P\mathrm{_{rf}^{out}}$ shown in b for $B$=0~T. b) Resonator model including a coupling capacitor ($C_\mathrm{c}$), followed by a resistance ($R_\mathrm{D}$), inductor ($L$) and a capacitance ($C_0=C_\mathrm{p}+C_\mathrm{D}$) in parallel. c) Measured reflection coefficient and fit before (grey) and after (purple/black) removing the offset angle $\phi$. The point of resonant frequency is marked as $f_\mathrm{r}$, whereas the off-resonant point corresponding to $f\rightarrow \infty$ is marked as $P'$. d) Absolute value of the reflection coefficient against the frequency, showing that the minimum of the raw data, does not correspond to the resonant frequency. e) Absolute value of the reflection coefficient variation when the SEB is at a charge instability at different frequencies. The maximum variation and, therefore SNR, occurs at $f_\mathrm{r}$. f) Phase of the reflection coefficient with respect to the frequency.  g), h) Absolute value and phase of the reflection coefficient on top of an SEB charge transition (top) and out of it (bg) and its respective resonant frequencies shown as vertical lines in their corresponding color. $|\Gamma|$ remains mostly the same, whereas the resonant frequency changes by 70kHz, revelling that the SEB impedance shift at a charge instability is mostly capacitive.} 
    \label{fig:Sup_smith}
\end{figure}

Here, we show data of our resonator at $B=0$~T in Fig.~\ref{fig:Sup_smith}a, whose equivalent model is presented in Fig.~\ref{fig:Sup_smith}b. A coupling capacitance $C_\mathrm{c}$ connects the transmission line to a parallel configuration of an inductor, $L$,  a resistor, $R_\mathrm{D}$, representing resonator losses and a variable capacitance $C_\mathrm{0}=C_\mathrm{p}+C_\mathrm{D}$, where $C_\mathrm{D}$ is the SEB capacitance from its gate and $C_\mathrm{p}$ is the parasitic capacitance of the circuit. The equivalent impedance of such resonator is given by~\cite{Ahmed2018}:

\begin{equation}
   Z_L=R_\mathrm{D}\frac{j\omega\Delta\omega_0}{\omega_0^2-\omega^2+j\omega\Delta\omega_0}+\frac{1}{j\omega C_c},
   \label{Eq:Z_L}
\end{equation}
where $\Delta\omega_0=\frac{1}{R_\mathrm{D} C_0}$ and $\omega_0=2\pi f_0=\frac{1}{\sqrt{{LC}_0}}$ are the width and resonant frequency of the unloaded parallel $LR_\mathrm{D}C_0$ circuit.

Assuming that $R_\mathrm{D}$ does not vary over the range of frequency near the resonance, the real part of the resonator's impedance stays constant, whereas the imaginary part (admittance) passes along different values. Such a behaviour corresponds to part of a circle in the complex plane when plotting the reflection coefficient $\Gamma=\frac{Z_L - Z_0}{Z_L + Z_0}$ at frequencies close to the resonance. The whole circle, so-called resistance circle, has its center on the real axis $\Im\left\{\Gamma\right\}=0$ and crosses it twice: firstly at the resonant frequency, $f_\mathrm{r}$, corresponding to the circle's closest point to the origin and a second time when the frequency tends to infinity and zero: $f\rightarrow \infty$ and $f\rightarrow 0$, at which $\Gamma=1$. For an example, see purple scatter circle in Fig.~\ref{fig:Sup_smith}c. 

In the case of perfect matching, $\Re\{Z_L\}=Z_0$, the circle is centered at the position $\Gamma=0.5$ and crosses the origin, leading to $\Gamma=0$ at $f_\mathrm{r}$. If the resonator is overcoupled ($\Re\{Z_L\}<Z_0=50$), the center of the circle is nearer the origin, making its radius larger than 1. In this case, the magnitude of the reflection coefficient $|\Gamma|$ does not tend to zero at the resonant frequency, but it's phase, $\angle{\Gamma}$, completes a whole $2\pi$ rotation as the frequency is varied across $f_\mathrm{r}$.  (See purple scatter in Fig.~\ref{fig:Sup_smith}c, d and e for an example of an overcoupled resonator).

A resistance circle in the complex plane  follows a complex Lorentzian scaled and transported from its origin~\cite{Pozar2005, Petersan1998}:

\begin{equation}
   \Gamma=o_c+\left(1-\frac{2}{1+j2Q_L(\frac{f}{f_\mathrm{r}}-1)}\right)r_\mathrm{c}.
\end{equation}


Here $o_c=\frac{r_\mathrm{L}}{1+r_\mathrm{L}}$ is the center of the circle, $r_c=\frac{1}{1+r_\mathrm{L}}$ is the radius, $r_\mathrm{L}=\frac{1}{\beta}$ is the real part of the normalised resonator impedance $\left(\frac{Z_L}{Z_0}=r_\mathrm{L}+jy_L\right)$, $f_\mathrm{r}$ is the resonant frequency and $Q_\mathrm{L}$ is the loaded quality factor defined as the ratio of the total energy stored in the resonator to the average energy dissipated per cycle multiplied by $2\pi$.  If we transport the circle centre to the origin, the change in phase is related to the frequency as

\begin{equation}
   \phi (\omega)=\theta_0+ 2\arctan{\left[2Q_\mathrm{L}(1-\frac{f}{f_\mathrm{r}})\right]},
\label{Eq:fit_phase}
\end{equation}

where $\theta_0$ is an offset angle. This is considered one of the most accurate ways to obtain the Q-factor and resonant frequency of a resonator~\cite{Petersan1998}.


Experimentally, the reflection coefficient is extracted by measuring the S-parameter $S_{21}$ between the lines driving $P\mathrm{_{rf}^{in}}$ and $P\mathrm{_{rf}^{out}}$ with a network analyzer as $\Gamma=\sqrt{S_{21}}=\sqrt{\frac{P\mathrm{_{rf}^{out}}}{ P\mathrm{_{rf}^{in}}}}$ (see Fig.~\ref{fig:Sup_smith}a). This measurement differs from the expected constant resistance circle due to the effect of the environment leading to~\cite{Probst2015}: 

\begin{equation}
    S_{21}=ae^{j\alpha}e^{-2  \pi jf \tau} \left( 1- \frac{Q_\mathrm{L}/|Q_\mathrm{e}|e^{j\phi}}{1+2iQ_\mathrm{L}(f/f_\mathrm{r}-1)} \right).
\end{equation}

Here, the constant $a$ takes into account that the amplitude of the outcoming wave has been modified by the attenuators and amplifiers present in the system. Moreover, due to the cable length, the wave has an electrical length characterised by $e^{j\alpha}$ and it  acquires a delay, $\tau$, that makes the phase proportionally dependent on the frequency as $e^{-j2\pi f\tau}$~\cite{Probst2015}.

Fig.~\ref{fig:Sup_smith}c shows in grey the resonator measured at $B$=0~T once the effect of the environment has been removed. The additional phase offset, $\phi$, is what produces an asymmetry in the absolute value of the reflection coefficient (See Fig.\ref{fig:Sup_smith}d). Only when the resistance circle is rotated to it's right position, the resonant frequency coincides with the minimum in the absolute value of the reflection coefficient (see purple circle in Fig.~\ref{fig:Sup_smith}d). The term $e^{j\phi}$ comes from  asymmetries of the resonator’s transmission  signal  due to  different  input  and  output impedances  at  the  two  ports  of  the  resonator~\cite{Khalil2012} or from standing waves in the transmission line connected to the resonator~\cite{Deng2012}. In order to fit the data to a circle and extract its center, and radius, $f_\mathrm{r}$ and $Q_\mathrm{L}$ we use a code based on the resonator tools python library found in \cite{resonaotr_tools} .

Figure~\ref{fig:Sup_smith}g and h show the magnitude and phase of the reflection coefficient with respect to the frequency at the top of an SEB charge instability (black) and out of it (purple). We found that the resonant frequency is $f\mathrm{_{r}^{top}}=686.099\pm 0.017$ MHz at the charge transition degeneracy point and $f\mathrm{_{r}^{bg}}=686.168\pm 0.016$ MHz away from it. The change in resonant frequency is linked to an increment in the SEB capacitance to ground as $f_\mathrm{r}=\frac{1}{2\pi\sqrt{L(C_\mathrm{c}+C_\mathrm{p}+C_\mathrm{D})}}$, being the change in capacitance $\Delta C_\mathrm{D}=0.09 \pm 0.03$ fF.

We observed that the system is overcoupled as the phase completes a $2\pi$ rotation but the circle does not cross the origin. The matching, calculated as $\beta=\frac{R}{Z_0}>1$, is barely changed by the SEB, being $\beta^\mathrm{top}=2.064$, $\beta^\mathrm{bg}=2.061$ at and away from the charge degeneracy point, respectively, and, neither is the loaded Q-factor: $Q\mathrm{_{L}^{top}}=125.3\pm0.5$ and
$Q\mathrm{_{L}^{bg}}=125.7\pm0.5$. This means that the charge instability in the SEB produces a capacitive change, which we confirm later. Because we're measuring a small change in the device capacitance, the maximum change in $|\Delta\Gamma|$ and, therefore SNR, occurs at the resonant frequency~\cite{Ahmed2018}, where the slope in phase is maximum (See Fig.~\ref{fig:Sup_smith}e).

As magnetic field is applied, the kinetic inductance varies, modifying the resonant frequency, Q-factor and matching. This way, the resonant frequency at $B$=2~T, at which spin-readout measurements were taken, is $f_\mathrm{r}=665.5$~MHz, the internal Q-factor is reduced to $Q_0=267$ and the matching is equal to $\beta=2.5$, as depicted in Fig.~1a and summarised Table 1 from the main text. 

\section{JPA calibration}
\label{SupSec:JPA}

\begin{figure}
\centering
\includegraphics[width=1\linewidth]{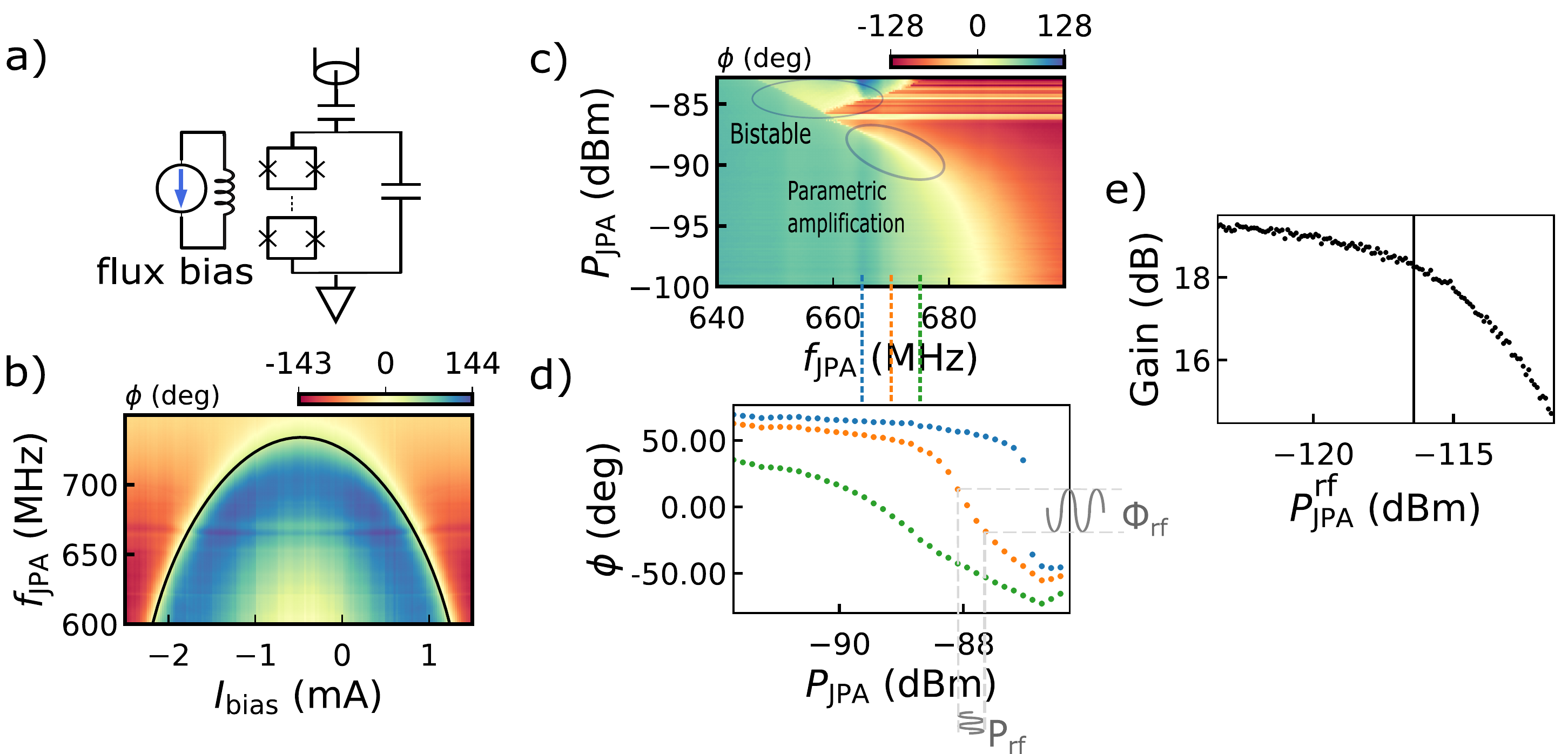}
\caption{a) Sketch of the JPA as a SQUID loop array in parallel with a shunted capacitance connected in reflection to the setup by a coupling capacitance. The magnetic flux generated in a nearby coil by the the current, $I_\mathrm{bias}$, is used to tune the JPA's resonant frequency. b) Reflected phase as a function of the pump frequency and $I_\mathrm{bias}$. The resonant frequency is fitted to a function proportional to $\mathrm{cosh}^2(I_\mathrm{bias})$. c) Reflected phase with respect to the pump frequency and the power applied to the JPA. In the region suitable for parametric amplification, the JPA's resonant frequency decreases as $P_\mathrm{JPA}$ is increased. d) Amplification transfer function at 3 different frequencies (663~MHz in blue, 668~MHz in orange and 673~MHz in green). A small variation in the power arriving to the JPA leads to a large variation of the reflected phase, producing a gain. e) Gain as a function of the rf-probe power arriving to the JPA, for a pump tone of frequency $f_\mathrm{JPA}=665.2$~MHz and power -88~dB. The back vertical line at -116~dBm denotes the rf-power for a 1dB compression in gain.}. 
    	\label{sup:JPA_tuning}
        \end{figure}

The JPA used in this experiment consists of a SQUID loop array shunted by a fixed capacitance, $C_\mathrm{JPA}$~\cite{Vijay2009}. This configuration creates a low quality factor ($Q_\mathrm{JPA}<100$) superconducting resonator, whose resonant frequency, $f\mathrm{_r^{JPA}}$, can be tuned from 550-750MHz (See Fig.~\ref{sup:JPA_tuning}b) by passing a current, $I_\mathrm{bias}$, through a nearby coil that modifies the flux through the SQUIDs.

In parametric amplification, one parameter is varied harmonically in a non-linear medium to create gain. The energy used to modulate the parameter is called the pump. In the case of the JPA, the non-linearity comes from the Josephson junction inductances, $L_\mathrm{J}$, that are varied harmonically when applying some power $P\mathrm{_{JPA}}$ at frequency $f\mathrm{_{JPA}}$ (see Fig.~1 of main text). Modifying $L_\mathrm{J}$, leads to changes in the JPA's resonant frequency, since $f\mathrm{_r^{JPA}}=\frac{2\pi}{\sqrt{C_\mathrm{JPA} L_\mathrm{J}}}$. Fig.~\ref{sup:JPA_tuning}c shows the variation of $f\mathrm{_r^{JPA}}$ as a function of $P_\mathrm{JPA}$. As $P_\mathrm{JPA}$ increases, $f\mathrm{_r^{JPA}}$ is firstly constant, but then it shifts to lower frequencies. Parametric amplification can be achieved in the power range in which  $f\mathrm{_r^{JPA}}$ varies with respect to $P_\mathrm{JPA}$. The JPA amplification transfer function is exemplified in  Fig.~\ref{sup:JPA_tuning}d, where small variations of the power arriving to the JPA due to the signal tone, $f_{rf}$, are translated into large changes in the reflected phase. When the JPA is tuned at even higher pump powers, it becomes bistable~\cite{Vijay2009} .

Figure~\ref{sup:JPA_tuning}e shows how, increasing the power of the signal tone arriving to the JPA, $P\mathrm{_{JPA}^{rf}}=P_\mathrm{rf}|\Gamma|^2$, leads to a gain reduction, since there is not enough pump energy to be transferred from the pump to the signal and idler. The power at which the gain is compressed by 1dBm is the JPA's dynamic range (-116~dBm).

\section{Lever arm and SEB to reservoir tunneling rates}
\label{sup:tun_rate_SEB}
 \begin{figure}
    \centering
    \includegraphics[width=0.4\linewidth]{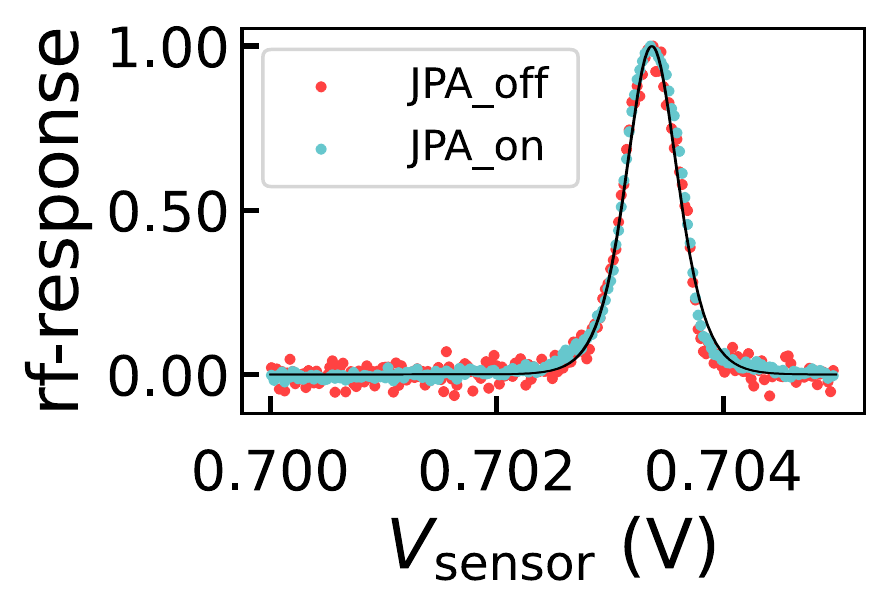}
    \caption{ Normalised  rf response from a SEB as a function of the voltage applied to its gate obtained using a low rf-tone power, $P_\mathrm{rf}=-91$~dBm. The width of the transition is related to the SEB-to-reservoir tunneling rate, leading to an upper limit of $\gamma\leq 74\pm12$~GHz. The same result is obtained with and without a JPA.} 
    \label{fig:supp_Te_JPA_ON_OFF}
\end{figure}   

The lever arm of $\alpha_\mathrm{S}=0.35\pm0.06$ is obtained using the slopes of the Coulomb diamonds measured in current~\cite{Hanson2007}.  On the other hand, $\gamma$ can be extracted using the  rf response of a SEB electronic transition, which is related to the SEB capacitance, $\Delta C_\mathrm{D}$ (See Fig.~\ref{fig:supp_Te_JPA_ON_OFF}). In the case that $k_\mathrm{B}T<h\gamma$, $\Delta C_\mathrm{D}$ depends on the electrochemical potential, $\epsilon$, as $\Delta C_\mathrm{D}\propto \frac{h\gamma}{\epsilon^2+(h\gamma)^2}$~\cite{Ahmed2018}. Knowing that $\epsilon=\alpha_\mathrm{S}V_\mathrm{S}$, we obtain $\gamma= 74\pm12$~GHz. 


 
\section{SNR calculation}
\label{SupSec_SNR}

\begin{figure}
    \centering
    \includegraphics[width=1\linewidth]{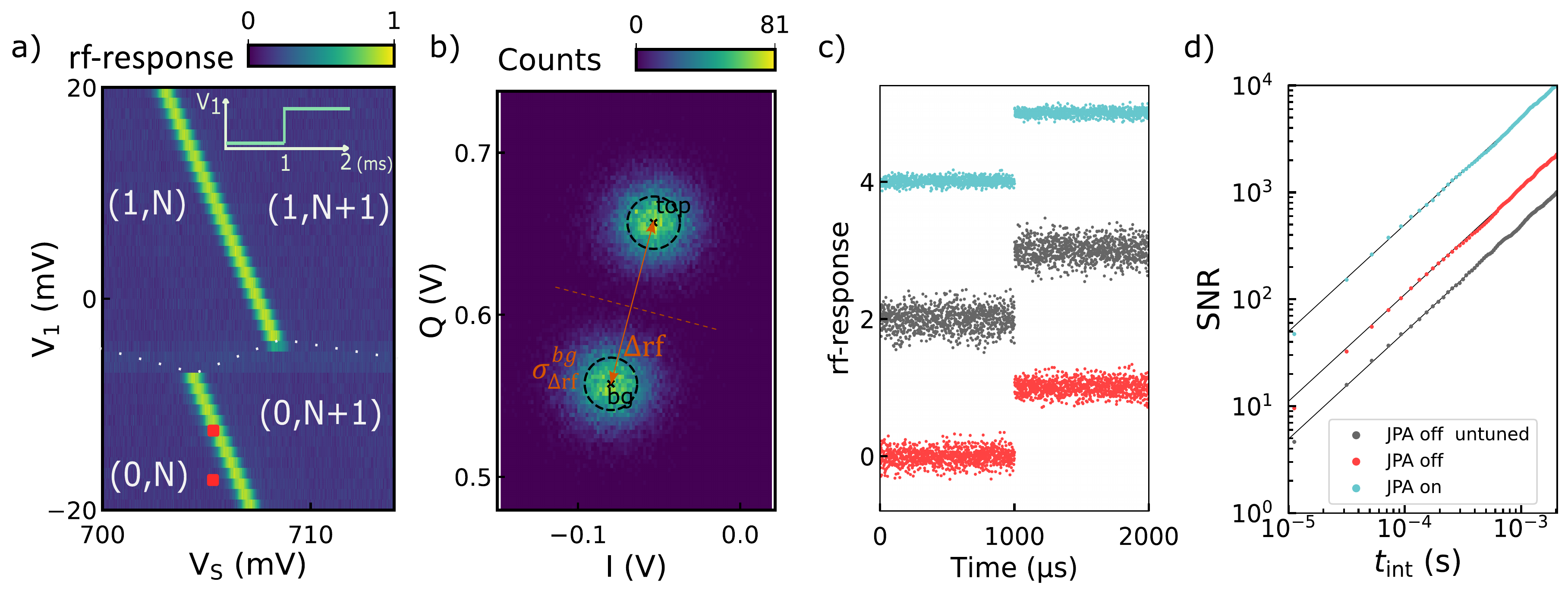}
    \caption{ SNR. a) Normalised  rf response showing the stability diagram of
the SEB versus $dot$ where the occupation of the SEB and $dot$ is displayed as ($dot$, SEB). Due to their cross capacitance, the  rf response has a shift in voltage when an electron is added to $dot$. The inset shows the pulses sent to $dot$  to jump on and off the dot-to-reservoir transition, which corresponds with the red points. b) I-Q histogram from a 1,000 data traces collected by pulsing between the red points marked in a). The histogram shows 2 distinct distributions corresponding to the background and the top of the SEB charge instability for data taken without a JPA. The signal is collapsed into 1D using the axis between the center of the so-called Fresnel lollipops. c) Normalised  rf response in the 1D-projection for JPA on, off and optimised JPA off for a measurement bandwidth $f_\mathrm{eff,BW}=12$~kHz. d) SNR as a function of the integration time.  }
    \label{fig:Sup_SNR_fig}
\end{figure}

To evaluate the readout performance, we send a 2-level pulse that varies the I-Q response between the top of the dot to reservoir transition (DRT) and the background (marked with red dots in Fig.~\ref{fig:Sup_SNR_fig}a). A histogram of the pulse  rf response in the quadrature plane shows two separated circular distributions, each one corresponding to the top and background of the DRT  (See Fig.~\ref{fig:Sup_SNR_fig}b). 

Since the noise is Gaussian and equal in every direction, most of the information is in the axis that joins the centers of the so-called Fresnel lollipops, whereas it's perpendicular axis carries just noise. Therefore, we project our data on the optimal axis and use the SNR definition $\mathrm{SNR}=\frac{\Delta \mathrm{rf}^2}{(\sigma_0^2+\sigma_1^2)/2}$, where $\Delta \mathrm{rf}$ is the distance between the lollipop centers and and $\sigma_{0(1)}$ is the 1-dimensional standard deviation of the background(peak).

Figure~\ref{fig:Sup_SNR_fig}d shows SNR as a function of integration time with the JPA off tuned at its optimal point (red), JPA on (blue) and JPA off with the same settings used for the JPA on (grey). Using an extrapolation (black straight lines) we infer the integration time to have an SNR=1. These times are $\tau_\mathrm{m}^\mathrm{off\, tuned}=902.1\pm0.1\, \mathrm{ns}$, $\tau_\mathrm{m}^\mathrm{off}=2.031\pm0.001\,\mathrm{\mu s}$, $\tau_\mathrm{m}^\mathrm{on}=200.7\pm0.8\, \mathrm{ns}$. This way, the noise temperature is reduced by a factor of x10 when switching the JPA on. However, the frequency at which this is achieved is not the optimal frequency, i.e. the natural frequency of the oscillator. The SNR when the JPA is off can be improved by a factor of x4.5 by choosing the optimal $f_\mathrm{rf}$ as it's showed in Fig.~1 from the main text. This is a consequence of the higher reflected power at $f_\mathrm{r}$ that partially saturates the JPA reducing its gain. 

\section{Electron temperature and thermal excitations}\label{SupSec_reservoir} 

     \begin{figure}
        \centering
        \includegraphics[width=1\linewidth]{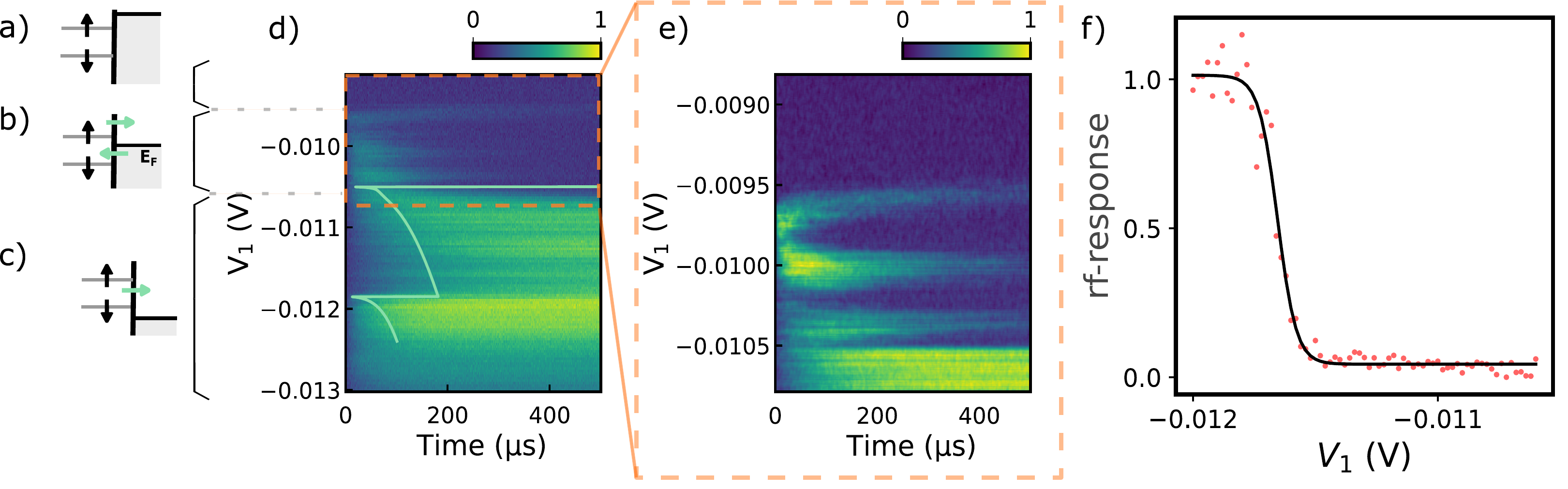}
        \caption{ a,b and c) show diagrams of $dot$ electrochemical potential with respect to the lead Fermi energy at different voltages applied to the $dot$ gate, $V_\mathrm{1}$. d) Time-averaged and normalised  rf response over time at different $V_\mathrm{1}$. At the voltages described by the situation in c), any loaded electron tunnels out during the readout stage. This corresponds with a low value of the  rf response that rise on time as the electron leaves $dot$. The superposed green line is the calculated tunneling times from a quasi-1D reservoir to a 0D dot. At higher $V_\mathrm{1}$, we encounter the situation pictured in b), where the reservoir Fermi energy is in between the dot spin $\ket{\uparrow}$ and $\ket{\downarrow}$. This regime is shown with more detail in e), where we can observe how the quasi 1D density of states from the lead shape the dependency of $t_\mathrm{out}^\uparrow$ and $t_\mathrm{in}^\downarrow$ with respect to $V_\mathrm{1}$. f)  rf response after the transient with respect to $V_1$, following a Fermi-Dirac distribution.} 
        \label{fig:supp_t_out}
    \end{figure}

Figure~\ref{fig:supp_t_out}a,b and c show different situations depending on the voltage applied to $dot$ at the readout stage, $V_1$. At high voltages, the electron stays in $dot$ (Fig.~\ref{fig:supp_t_out}a), corresponding to a constant rf response equal to zero, as shown in Fig.~\ref{fig:supp_t_out}d. As the voltage is decreased, the $\ket{\uparrow}$ and $\ket{\downarrow}$ states straddle
the reservoir Fermi energy, leading to the transitory behaviour in the rf response that allow us to determine whether the electron had a spin-up or spin-down (See Fig.~\ref{fig:supp_t_out}e).

At even lower voltages, the electron leaves the dot during the readout stage independently of it's spin polarisation (Fig.~\ref{fig:supp_t_out}e).  At this voltage range, we observe the characteristic features of tunneling between a 0-dimensional and a 1-dimensional system,  where the tunneling rate depends on the energy as $\Gamma \propto \frac{1}{\sqrt{ E-E_\mathrm{n}}}$, being $E_\mathrm{n}$ the position in energy of the reservoir 1D subbands . The simulated tunneling time as a function of $V_1$ is superposed over the 2D-map in Fig.~\ref{fig:supp_t_out}d with a green line. 

Around the Fermi level, the reservoir density of states follow a Fermi-Dirac distribution, so that the tunnling rate is a combination of the 1-D subbands and the Fermi-Dirac distribution:

\begin{equation}
   \Gamma(E) = \frac{2\pi}{\hbar}\vert \Gamma_0\vert \left(\sum_n \frac{1}{\sqrt{ E-E_\mathrm{n}}}\right) (1-f(E-E_\downarrow)).
\end{equation}

Here $E=-|e|\alpha_1 V_1$, where $\alpha_1$ is the lever arm of the gate over $dot$ and $e$ is the electronic charge. $1-f(E-E_\downarrow)$ is the distribution of empty states in $dot$, which is tracked by the rf response after the transient tunneling. The electronic temperature, $T_\text{e}=137\pm 18$~mK, is extracted by fitting the rf response as a function of $V_1$ to the Fermi-Dirac distribution (See Fig.~\ref{fig:supp_t_out}f).

We use the calculated Fermi distribution to obtain the thermal excitation time constant, $t_{out}^\downarrow$, at the readout position as $t_\mathrm{out}^\downarrow =t_\mathrm{in}^\downarrow \frac{(1-f(E_\mathrm{readout}-E_\downarrow )}{f(E_\mathrm{readout}-E_\downarrow)}$ (See \S\ref{SupSec_parameter} for more information about $t_\mathrm{in}^\downarrow$). We obtain a $t_\mathrm{out}^\downarrow=309$~s and $t_\mathrm{out,JPA}^\downarrow=70$~s for measurements taken without and with a JPA, respectively.

\section{Experimental bandwidth}
\label{sup:experimental_bandwidth}
The bandwidth of our experiment is limited by the resonator bandwidth: $\frac{f_\mathrm{r}}{Q_\mathrm{L}}=6.18\pm0.04$ MHz. However, if the signal's frequency components are lower than the resonator bandwidth, a low-pass filter can be introduced to reduce high frequency noise, improving the SNR.

To characterise the measurement bandwidth, we can obtain the effective noise bandwidth as

\begin{equation}
   \omega_\mathrm{eff,BW}=\int_{0}^{\infty} \left|\frac{H(j\omega)}{H_\mathrm{max}}\right|^2 \, d\omega,
\end{equation}

which corresponds to the bandwidth of a brickwall filter that produces the same integrated noise power. Here, $H(j\omega)$ is the filter transfer function, and $H_\mathrm{max}$ is its maximum. 

In this experiment, we used a (minicircuits BLP-1.9+) low pass filter, whose transfer function was obtained from its insertion loss provided by the manufactured as
\begin{equation}
 \mbox{Insertion loss (dB)}=10\log_{10}\left|\frac{V_\mathrm{i}}{V_\mathrm{f}}\right|^2=-20\log_{10}|H_\mathrm{MC}(j\omega)|,
\end{equation}

where $V_\mathrm{i}$ and and $V_\mathrm{f}$ ire the filter's input and output voltage, respectively.

\begin{figure}
\centering
\includegraphics[width=1\linewidth]{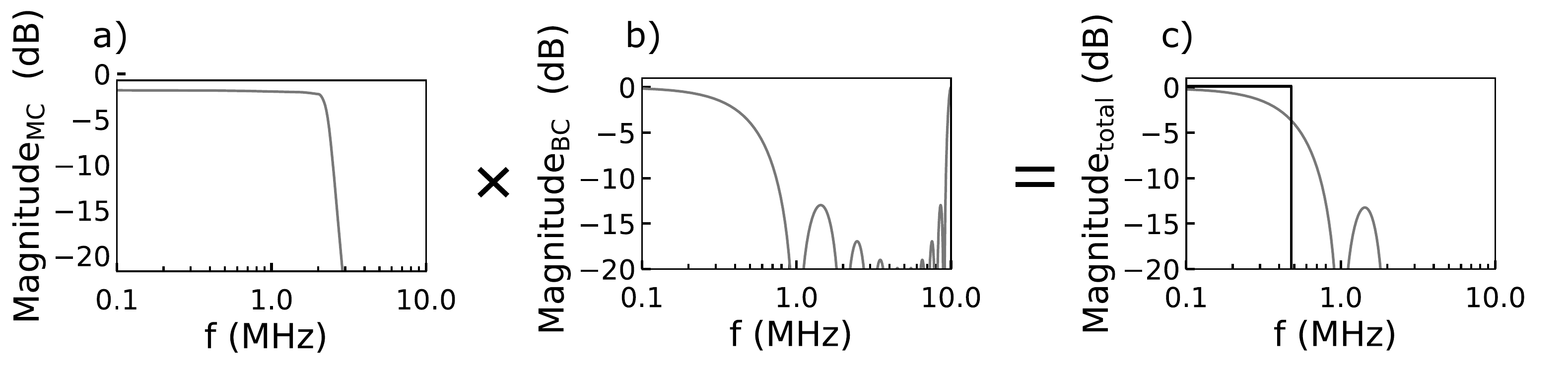}
\caption{a) Bode diagram of the magnitude of the 8th-order minicircuits BLP-1.9+ filter utilised in our measurements. b) Same for the boxcar filter used to downsample the sample rate from 10~MHz to 1~MHz. c) Combination of the effects from both filters and equivalent brickwall filter with the same integrated noise power, showing an effective noise bandwidth of  $f_\mathrm{eff,BW}=0.49$~MHz. } 
    	\label{fig:supp_t_int}
        \end{figure}

After that filter, a digital boxcar filter that averages every ten points ($N$=10) is applied, followed by a decimation process, to reduce the sample rate from 10~MHz to 1~MHz. This way, the total transfer function is equal to $|H_\mathrm{total}(j\omega)|=|H_\mathrm{MC}(j\omega) H_\mathrm{BC}(j\omega)|$, leading to the filter depicted in Fig.~\ref{fig:supp_t_int}, with an effective noise bandwidth of $f_\mathrm{eff,BW}=\frac{\omega_\mathrm{eff,BW}}{2\pi}=0.49$ MHz.

On top of that, some of our measurements include a rolling average filter that takes the average of every $N$ points recursively. This additional filter modifies the total transfer function as $|H_\mathrm{total}(j\omega)|=|H_\mathrm{MC}(j\omega) H_\mathrm{BC}(j\omega)H_\mathrm{RF}(j\omega)|$, where $H_\mathrm{RF}(j\omega)$ is the rolling average transfer function (See \S\ref{SupSec_rolling} for more information).

\section{Parameter extraction}
\label{SupSec_parameter}
    

The electrical fidelity for spin-dependent measurements is calculated by simulating single-shot histograms as the ones shown in Fig.~2b and c. In order to create them, we need to reproduce single-shot traces equivalent to the ones measured.  An example of a spin $\ket{\uparrow}$ trace is depicted in  Fig.~~\ref{fig:supp_parameters_E}a with a blip starting at  $t_\mathrm{out}^\uparrow$, lasting for $t_\mathrm{in}^\downarrow$. The background and blip have values $E$(low) and $E$(high), with its respective noise, $\sigma_\mathrm{low}$ and $\sigma_\mathrm{high}$. This example can be labelled as a spin $\ket{\uparrow}$ trace since it surpasses the threshold voltage, $V_\mathrm{T}$.

The experimental parameters that ultimately determine those traces can be separated into the ones that depend on the sensor and the ones that depend on the measured dot. The sensor parameters are independent on the readout method and are the rf response at the background, $E$(low), and at the blip , $E$(high), and their respective noise level. 

\begin{figure}
    \centering
    \includegraphics[width=0.8\linewidth]{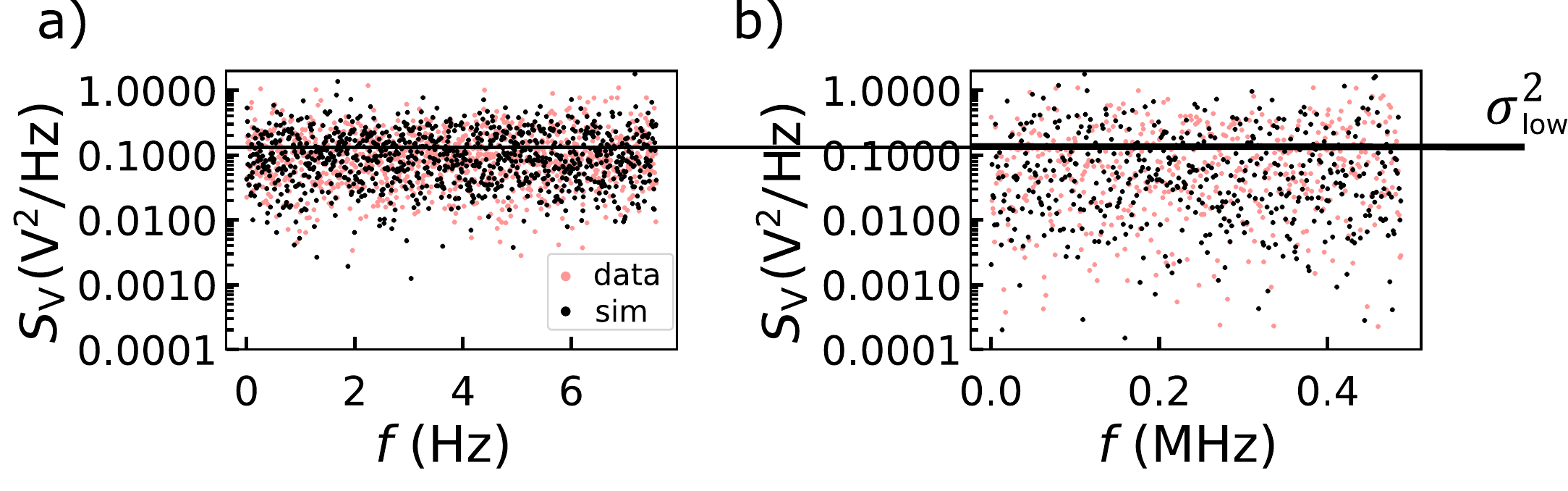}
    \caption{Noise spectral density of the  rf response at $E$(low) at low frequencies (a) and high frequencies (b) for the acquired data (pink) and the simulations created with Gaussian noise (black).} 
    \label{fig:supp_S_noise_gaussian}
    \end{figure}

In order to extract these parameters, 10,000 single-shot spin readout traces like the one displayed in Fig.~\ref{fig:supp_parameters_E}a were registered. The average of the rf response at the blip and at the background are equal to the expectation values $E$(high) and $E$(low), respectively. To characterise the noise level of the background, we obtain the noise spectral density, $S_\mathrm{V}(f)$ of the rf response (see Fig.~\ref{fig:supp_S_noise_gaussian}a and b). For lower frequencies, the background noise spectral density is obtained using the last data point of all the consecutive readout traces (see Fig.~\ref{fig:supp_S_noise_gaussian}a), whereas for higher frequencies we calculate $S_\mathrm{V}$ of a single spin $\ket{\downarrow}$ trace (see Fig.~\ref{fig:supp_S_noise_gaussian}b). Comparing the noise spectrum with the one generated by a Gaussian random number generator, we concluded that the noise of the background has a Gaussian profile with variance $\sigma^2_\mathrm{low}$ for the whole set of measurements. 

\begin{figure}
    \centering
    \includegraphics[width=0.8\linewidth]{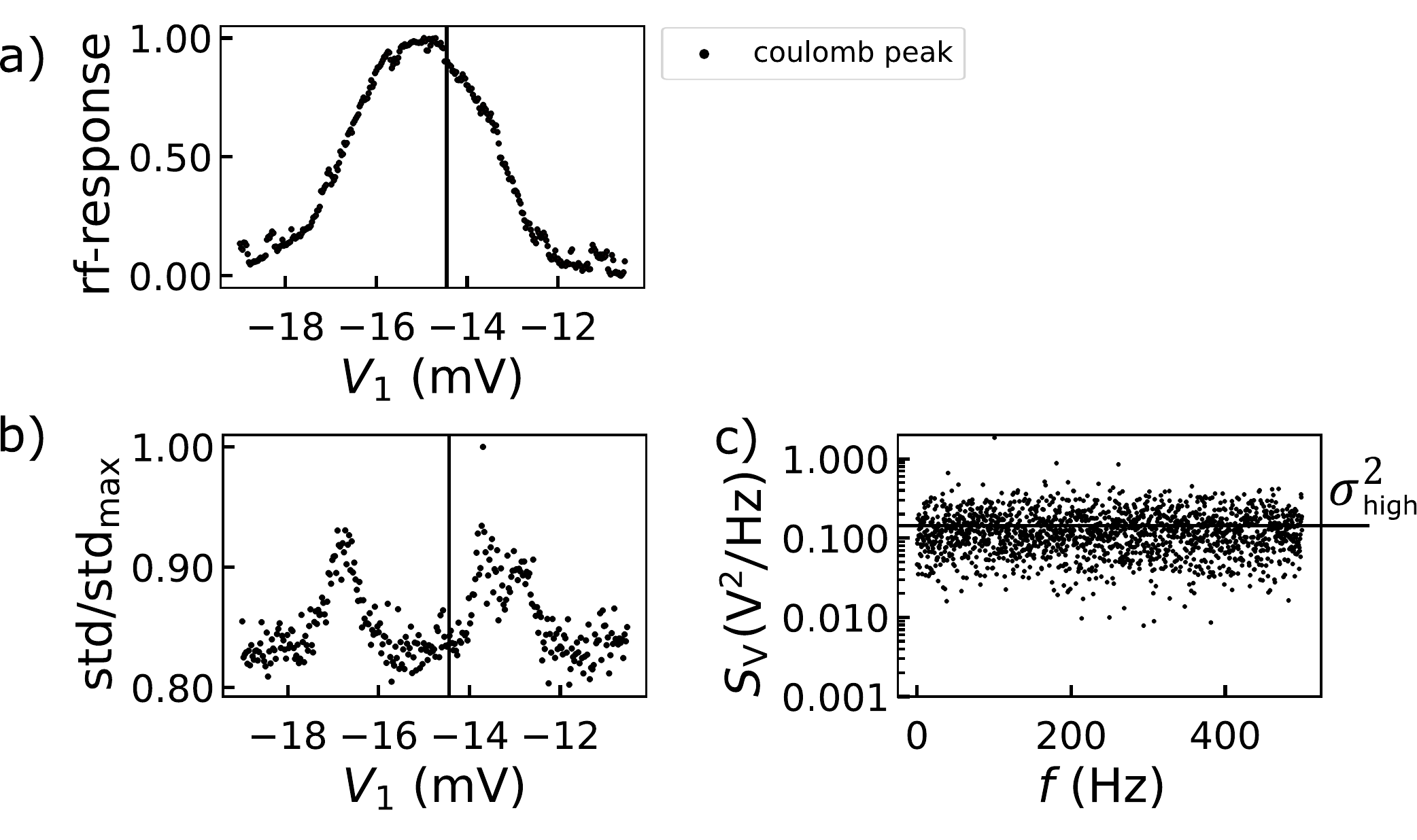}
    \caption{a) Average of the normalised  rf response for a SEB electronic transition with respect to the voltage applied to the $dot$ gate, $V_\mathrm{1}$. The vertical line indicates the voltage for spin readout. b)Standard deviation over 2~ms of rf response at $E$(high) at different voltages. c) $S_\text{V}$ at the blip.} 
    \label{fig:supp_noise_vs_B2}
    \end{figure}

\begin{figure}
    \centering
    \includegraphics[width=0.8\linewidth]{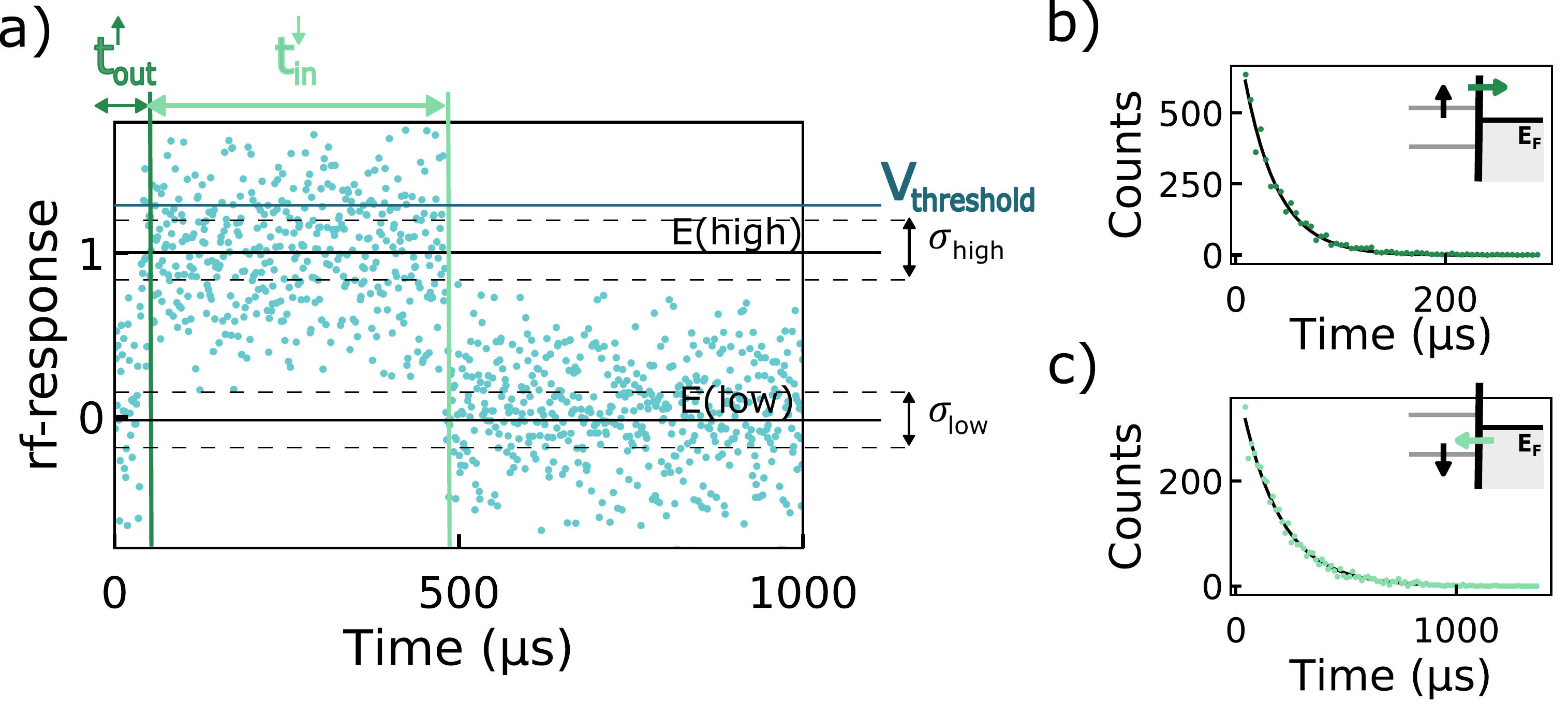}
    \caption{a) Normalised  rf response of a spin $\ket{\uparrow}$ data trace taken with a JPA using a sample rate of $f_s$=1~MHz and a measurement bandwidth of $f_\mathrm{eff, BW}=0.49$~MHz. The blip starts when the spin $\ket{\uparrow}$ electron leaves the dot at $t_\mathrm{out}^\uparrow$ and lasts until a spin $\ket{\downarrow}$ electron replaces it ($t_\mathrm{in}^\downarrow$ ). When $dot$ is occupied the  rf response has an estimated value $E$(low) with a standard deviation $\sigma_\mathrm{low}$, whereas when it's empty the estimated value and standard deviation are $E$(high) and $\sigma_\mathrm{high}$, respectively. We have also indicated the $V_\mathrm{threshold}$ above which the trace is labeled as a spin $\ket{\uparrow}$. b) Histogram of the starting point of the pulse and exponential fit. c) Histogram of the pulse duration and fit. }
    \label{fig:supp_parameters_E}
\end{figure}

The noise at $E$(high) can include can include additional sources of noise such as charge noise, where the noise spectral density typically depends on the frequency as $1/f$. This noise is originated from the collective behaviour of defects or charge traps that act as charge fluctuators as they trap and release electrons~\cite{Kranz2020}. The charge fluctuations slightly modify the potential around the sensor modifying its bias point, so that their effect is more noticeable at the slope of a SEB electronic transition than at the top. Figure ~\ref{fig:supp_noise_vs_B2}a shows the SEB dot-to-reservoir transition as a function of the voltage applied to $dot$, where the signal at each point has been averaged over 2 ms. Figure~\ref{fig:supp_noise_vs_B2}b displays their corresponding standard deviations. It's very clear how the standard deviation is higher on the slope than on the offset and top of the peak.

We observe that the readout position -- marked as a black vertical line -- is at the maximum of the  rf response. This has two benefits, on one hand, the contrast between $dot$ empty and occupied is maximum ($E$(low) and $E$(high)) and, on the other hand, the charge noise is minimised. Fig.~\ref{fig:supp_noise_vs_B2}c shows that the noise spectral density at the top of the  rf response is also constant over the range of frequencies with a variance $\sigma^2_\mathrm{high}$ very similar to $\sigma^2_\mathrm{low}$. 

The rest of parameters ($t_\mathrm{out}^\uparrow$, $t_\mathrm{in}^\downarrow$ and $A$) are set by $dot$. $t_\mathrm{out}^\uparrow$ is the time constant for a spin $\ket{\uparrow}$ electron to leave the dot. Such time corresponds with the start time of the blip and can be determined as the time at which the  rf response reaches certain threshold voltage, $V_\mathrm{T}$. Registering the number of times that the  rf response exceeds such threshold voltage at a given readout time follows an exponential trend whose time constant is $t_\mathrm{out}^\uparrow$. $t_\mathrm{in}^\downarrow$ is obtained following a similar analysis, where the blip duration probability is fitted  to an exponential function (See Fig.~\ref{fig:supp_parameters_E} b and c).  

The simulated traces for a spin $\ket{\downarrow}$ were created as a set of points with a sample rate of $\Gamma_\mathrm{s}=$1~MHz (as the one of the experiment) and constant value $E$(low) to which it's added a Gaussian noise characterised by $\sigma_\mathrm{low}$.  Spin $\ket{\uparrow}$ traces are generated as a constant value $E$(low)  with Gaussian noise $\sigma_\mathrm{low}$ and a blip with a constant value of $E$(high) and a standard deviation of $\sigma_\mathrm{high}$. The blip starting time and duration follow  exponential distributions with time constant $t_\mathrm{out}^\uparrow$ and $t_\mathrm{in}^\downarrow$, respectively.

\subsection{Moving average filter}
\label{SupSec_rolling}
The SNR can be increased in post-processing by adding an additional low-pass filter, which removes high frequency noise at the price of affecting the blip shape when the cut-off frequency is too low. From the many digital filters available, we used the rolling average filter which takes the average over $N$ points recursively. The first point of the filtered signal corresponds to the mean of the first $N$ points from the original signal and the subsequent points are obtained by shifting forwards by one time step the subset of $N$ points that is averaged.  

Although the rolling average has a complicated frequency dependence (See Fig.~\ref{fig:supp_t_int}b), it's ideal for this application since it has one of the lowest computation times and is optimal to reducing random noise while retaining a sharp step response~\cite{Smith_book}. This way, we can test which measurement bandwidth maximizes the readout fidelity with a low computation overhead. 
    
\section{Measurement fidelity}\label{SupSec_meas_fidelity}  
The probability of correctly recognising a spin $\ket{\uparrow}$ electron, $F_\mathrm{M}^\uparrow$,is given by the sum of the probability that such electron truthfully generates a spin $\ket{\uparrow}$, $F_\mathrm{STC}^\uparrow$, and the sensor detects the corresponding blip, $F_\mathrm{E}^\uparrow$,  plus the probability of generating a false  spin $\ket{\downarrow}$ trace, $1-F_\mathrm{STC}^\uparrow$,  that is misidentified as a spin $\ket{\uparrow}$ electron, $1-F_\mathrm{E}^\downarrow$. This way,

   \begin{equation}
     \label{Eq:F_M_UP}
        F_\mathrm{M}^\uparrow=F_\mathrm{STC}^\uparrow F_\mathrm{E}^\uparrow + (1-F_\mathrm{STC}^\uparrow)(1-F_\mathrm{E}^\downarrow).
    \end{equation}
    
Equivalently, the probability of correctly recognise a spin $\ket{\downarrow}$ electron is:
   \begin{equation}
     \label{Eq:F_M_DOWN}
        F_\mathrm{M}^\downarrow=F_\mathrm{STC}^\downarrow F_\mathrm{E}^\downarrow + (1-F_\mathrm{STC}^\downarrow)(1-F_\mathrm{E}^\uparrow).
    \end{equation}

Both independent fidelities can be combined as
   \begin{equation}
     \label{Eq:F_M}
        F_\mathrm{M}=\frac{F_\mathrm{M}^\downarrow+F_\mathrm{M}^\uparrow}{2}
    \end{equation}
    
to calculate the overall measurement fidelity. The electrical fidelity is calculated via a Monte-Carlo simulations as described in \S\ref{SupSec_parameter}, whereas $F_\mathrm{STC}$ fidelity uses an analytic expression to take into account the errors coming from relaxation and thermal processes. The probability of not having a thermal excitation, so a spin  $\ket{\downarrow}$ does not produce a false spin $\ket{\uparrow}$ trace, is given by:

\begin{equation}
     F_\mathrm{STC}^\downarrow={e^{-t/t_\mathrm{out}^\downarrow}}.
\end{equation}

The other infidelity source is the relaxation process of a spin $\ket{\uparrow}$ electron that hasn't tunneled out of the dot. That can be calculated as the conditional probability $P(A|B)$, being $P(A)=1-{e^{-t/T_\mathrm{1}}}$ the probability that an spin $\ket{\uparrow}$ has decayed at time $t$, and $P(B)=e^{-t/t_\mathrm{out}^\uparrow}$, the probability that an electron with spin $\ket{\uparrow}$ hasn't left the dot at time, $t$. Since both events are independent the probability of having a false spin $\ket{\downarrow}$ trace due to a relaxation process is:

\begin{equation}
    P(A|B)=\frac {P(A\cap B)}{P(B)}= \frac {P(A)P(B)}{P(B)}=P(A)=1-{e^{-t/T_\mathrm{1}}},
\end{equation}

Therefore, the probability of not having a relaxation process is $e^{-t/T_\mathrm{1}}$. To calculate the fidelity, we have to add the probability that an spin $\ket{\uparrow}$ relaxes and, subsequently the spin $\ket{\downarrow}$ electron tunnels down the dot:

\begin{equation}
    F_\mathrm{STC}^\uparrow={e^{-t/T_\mathrm{1}}}+(1-{e^{-t/T_\mathrm{1}}})(1-{e^{-t/t_\mathrm{out}^\downarrow}})
\end{equation}

Here, the relaxation time $T_\mathrm{1}$ depends on the magnetic field applied~\cite{Huang2014}, which in this experiment was $5.1\pm0.4$ s, at $B$=2~T. On the other hand, $t_\mathrm{out}^\downarrow$ depends on the temperature, and the difference in energy between the spin $\ket{\downarrow}$ and the reservoir Fermi Energy at the readout stage.
\section{$F_\mathrm{M}$ dependence on $\Delta t$ and measurement bandwidth}\label{SupSec_Deltat} 

\begin{figure}
        \centering
        \includegraphics[width=0.8\linewidth]{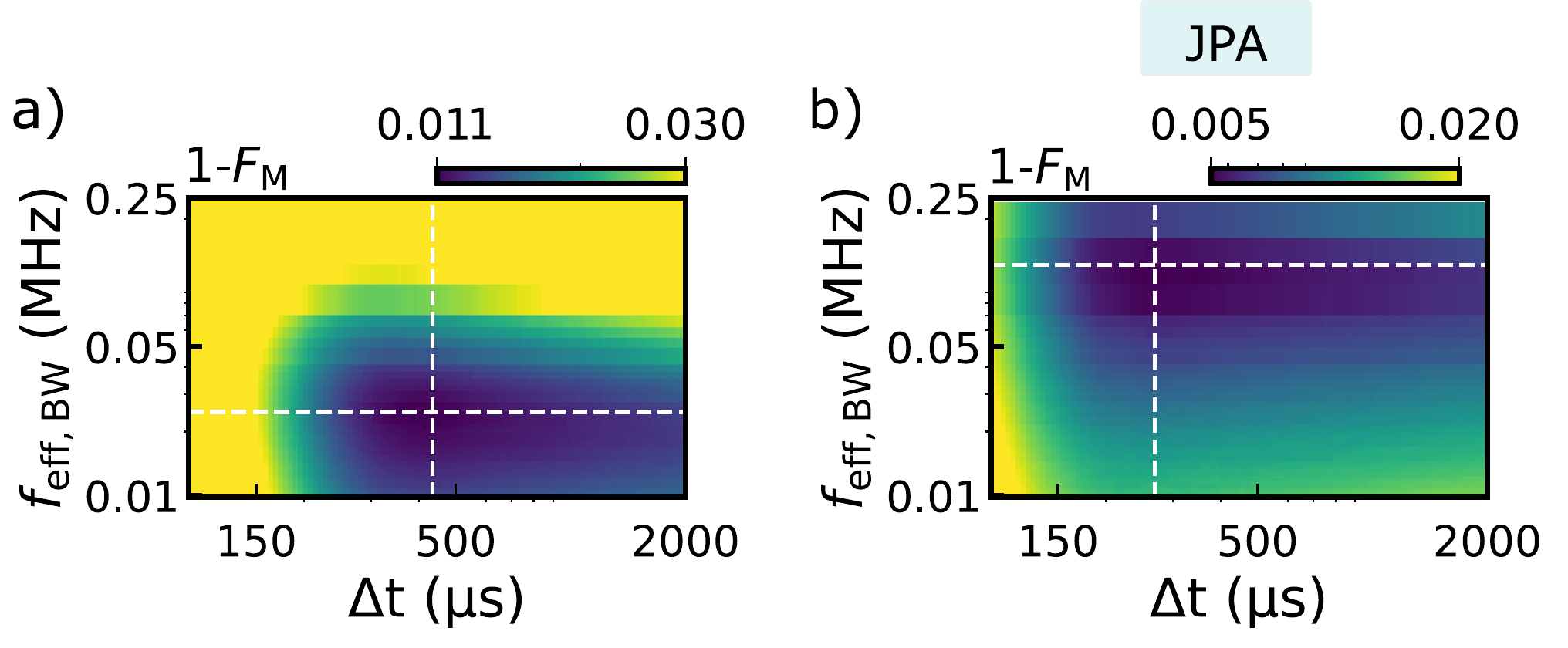}
        \caption{ a). Dependence of the measurement infidelity $1-F_\mathrm{M}$, with respect to the measurement bandwidth and the readout time. The dashed white lines pass through the maximum fidelity point. b) Same for measurements taken using  JPA.} 
        \label{fig:supp_FM_2D}
    \end{figure}

Here, we investigate the dependence of the measurement fidelity $F_\mathrm{M}$ with respect to the measurement bandwidth and the readout time with and without a JPA. On one hand, decreasing the measurement bandwidth improves the SNR, but, on the other hand, it deforms the blip shape, rounding its edges and decreasing its maximum. Figure~\ref{fig:supp_FM_2D}a and b show how decreasing the measurement bandwidth increases the fidelity up to an optimal point. The difference in the optimal measurement bandwidth is due to two different reasons: first the measurements taken without a JPA have a lower SNR. So that, the limiting factor to increase the fidelity is the noisy spin $\ket{\downarrow}$ traces reaching above the threshold.  The second reason is that the tunneling rates measured for each setup were slightly different: the blips have a standard duration of ${t_\mathrm{in}^\downarrow}=440~ \mathrm{\mu}s$ without a JPA and ${t_\mathrm{in, JPA}^\downarrow}=186~\mathrm{\mu}s$. So that, the optimal measurement bandwidth is higher for the set of measurements without a JPA.

$F_\mathrm{M}$ also increases with $\Delta t$, since more blips can be captured as the readout time duration is longer. However, once the readout time is longer that the standard duration of the spin dependent tunneling process, the rest of the trace can only lead to errors. For this reason, the optimal readout time with a JPA is shorter, having faster tunneling times ${t_\mathrm{out, JPA}^\uparrow}$ and ${t_\mathrm{in, JPA}^\downarrow}$. The white dashed lines in Fig~.\ref{fig:supp_FM_2D} passes through the maximum in $F_\mathrm{M}$ and correspond with the 1D-plots presented in Fig.~2c and d in the main text.



\section{Machine learning spin labelling approach}
\label{sec:machine_learning}

\begin{figure}
\centering
\includegraphics[width=0.7\linewidth]{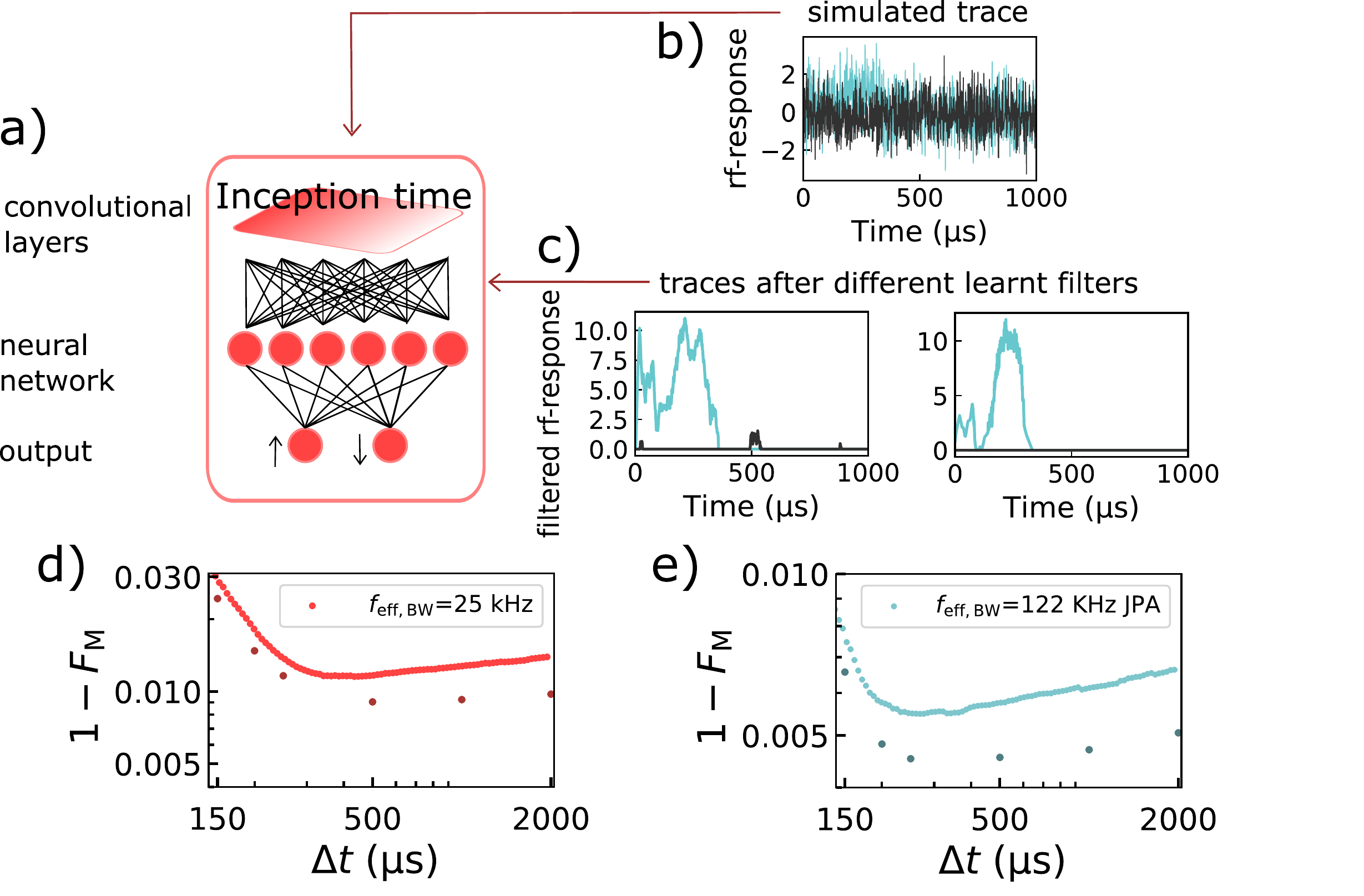}
\caption{a) Working protocol of the machine learning spin-classification approach. b) Simulated spin $\ket{\uparrow}$ and $\ket{\downarrow}$ trace using the parameters for JPA on. c) Output of the InceptionTime convolution layers with the blip edges enhanced to facilitate trace classification. d) Measurement fidelity without a JPA using the threshold method approach (light red) and the machine learning approach (dark red). d) Same for measurements using a JPA, with threshold (light blue) and machine learning (dark blue) labelling methods.} 
    	\label{fig:supp_ML}
        \end{figure}

The main text describes how to obtain the electrical fidelity using the probability density function of the  rf response peak values (Eq.~3). However, when applying other spin identification methods, the fidelity can be calculated with an equivalent method that relies on the number of simulated traces wrongly identified, being:

\begin{equation}
    F\mathrm{_E^\uparrow}=1-n^0_\uparrow/N_\mathrm{tot} \,\,\,\,\,\,\,\,\,\,\,\, F\mathrm{_E^\downarrow}=1-n^1_\downarrow/N_\mathrm{tot}.
\end{equation}

Here, $n^0_\uparrow$ is the number of spin $\ket{\uparrow}$ traces missidentified as $\ket{\downarrow}$ traces, and the opposite holds for $n^1_\downarrow$. We use this method to calculate the measurement fidelity when using a neural network method to label the readout traces.

The neural network method is summarised in Fig.~\ref{fig:supp_ML}a. It uses the deep learning architecture known as InceptionTime, a state of the art approach to time series classification. The InceptionTime network involves a series of convolutional layers which apply learned filters to the time series to extract its features~\cite{IsmailFawaz2020}. The features extracted from Fig.~\ref{fig:supp_ML}b spin traces are shown in Fig.~\ref{fig:supp_ML}c. These features are fed into a fully connected or dense layer which assigns one of two classes to the input time series (spin up or down). The network was trained using the same body of data that the thresholding method described in the main text, with it divided into training, validation, and test sets. The training set is used to train the model via gradient descent and the validation set is monitored during training to avoid overfitting. If the network learns to recognise the training set too well, then that can compromise its performance on unseen data. The network that produces the best validation accuracy is selected and is applied to the test set, which gives the final accuracy data reported here. The approach used here uses the TSAI package for instantiating the networks and records training metrics using the Weights and Biases library, which is also used for hyperparameter optimisation~\cite{tsai, wandb}.

Figure~\ref{fig:supp_ML}d and e show an improvement of the fidelity for measurements taken with and without a JPA. Here, the spin to charge errors are also included as described in \S\ref{SupSec_meas_fidelity}. We find a maximum fidelity $F_\mathrm{M}=99.1\%$  for $\Delta t=500$~$\mu$s without using a JPA and $F^\mathrm{JPA}_\mathrm{M}=99.54\%$ with a JPA for $\Delta t_\mathrm{JPA}=250$~$\mu$s. We observed that, when using the machine learning classification method,  the measurement fidelity stays almost constant  as the readout time, $\Delta t$, increases. This is because the  optimised filters enhance the blip's edges features, mitigating the errors that appeared in the threshold method when the background
noise surpasses the threshold.

\section{Measurement fidelity for asymmetric tunneling rates}
\label{sec:PSB_fid_simulation}

The readout time for spin-dependent tunneling is limited by  the time that a spin $\ket{\uparrow}$ electron takes to leave the dot, $t_\mathrm{out}^{\uparrow}$, since until the start of the blip there's no difference between a spin $\ket{\uparrow}$ and $\ket{\downarrow}$ trace. 
\begin{figure}
\centering
\includegraphics[width=1\linewidth]{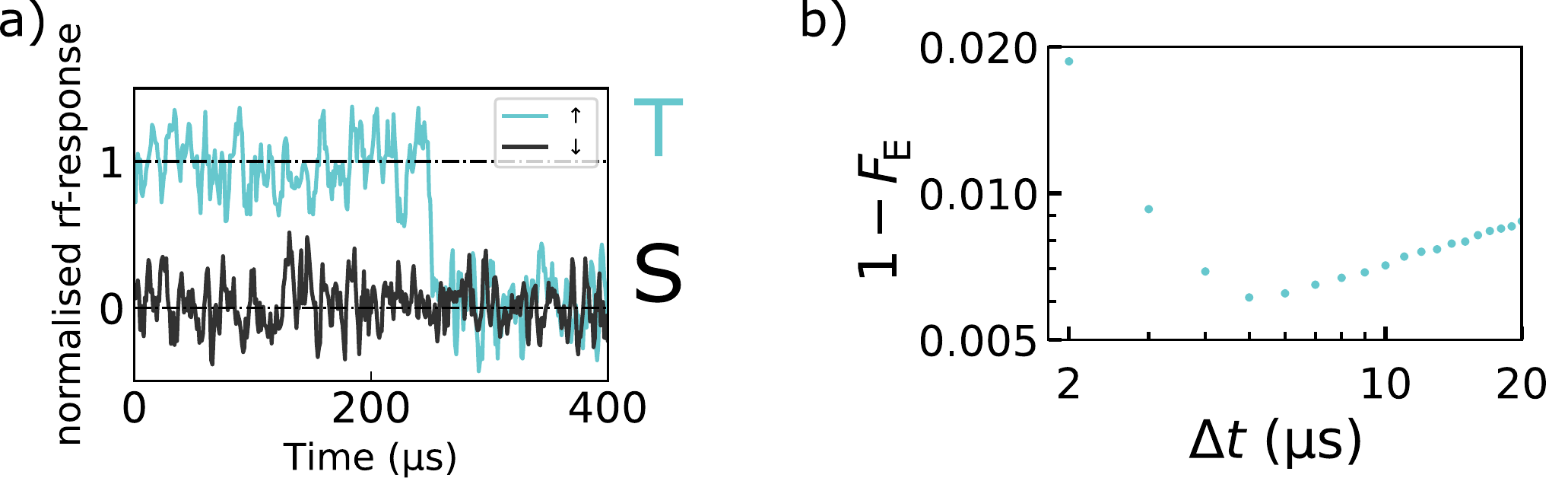}
\caption{a) Spin $\ket{\uparrow}$ and  $\ket{\downarrow}$ traces simulated using the experimental parameters of measurements taken with a JPA, with the exception of $t_\mathrm{out}^{\uparrow}$ and $t_\mathrm{in}^{\downarrow}$ that are modified to emulate triplets and singlets. The spin $\ket{\uparrow}$ is equivalent to a triple trace and, in the same way, the spin $\ket{\uparrow}$ is equivalent to a singlet trace. The traces shown have a readout bandwidth of $f_\mathrm{eff,BW}=122$ kHz.b) $1-F_\mathrm{E}$ as a function of the readout time.} 
    	\label{fig:sup_PSB_sim}
        \end{figure}

Here, we investigate the measurement fidelity for asymetric tunneling rates -- a fast $t_\mathrm{out}^{\uparrow}$, while $t_\mathrm{in}^{\downarrow}$ remains long -- to reduce the readout time necessary to achieve a fidelity above $99\%$. This kind of traces are shown in Fig.~\ref{fig:sup_PSB_sim}a and are very similar to the ones described in Fig.~3d from the main text, showing singlet and triplet traces. 

We can make a parallelism between the traces generated using spin-dependent tunneling and Pauli spin blockade (PSB). In both cases the value of the rf-response depends on the dot occupation. In the case of (PSB), the rf-response is maximum for (4,0) and minimum for (3,1), whereas for spin-dependent tunneling is maximum when the dot is empty and minimum when it has an electron. This way, both singlet, and spin $\ket{\downarrow}$ traces are characterised by a constant rf-response. On the other hand, a $\ket{\uparrow}$ trace has a blip that starts when the electron leaves the dot, $t_\mathrm{out}^{\uparrow}$, and lasts until a new electron tunnels back from the reservoir to the spin $\ket{\downarrow}$ state, $t_\mathrm{in}^{\downarrow}$. In the case of a triplet trace, the blip starts when the the system tunnels from the preparation stage, with an occupation of (4,0), to the readout stage, with occupation (3,1). Such a tunneling rate is can be fast that it's not registered in the measurement. Therefore, a triple trace starts at a high value that continues until the triplet relaxes to the singlet, characterised by the relaxation time, $T_1$.

To obtain the fidelity, we create traces in the same way described in \S\ref{SupSec_parameter} and using the same experimental values extracted for measurements taken with a JPA ($E$(high), $E$(low), $\sigma_\mathrm{high}$, $\sigma_\mathrm{low}$, $\Gamma_\text{s}$ and proportion of spin $\ket{\downarrow}$, $A$). However, the tunneling rates are modified in order to emulate triplet/singlet traces. We chose $t_\mathrm{out}^{\uparrow}=0.01$~$\mu$s and $t_\mathrm{in}^{\downarrow}=228$~$\mu$s, so that $t_\mathrm{in}^{\downarrow}$ is equal to the triplet relaxation time $T_1$ from the Pauli Spin blockade experiment described in the main text.

The average of the trace during $\Delta t$ is compared against a threshold, which is varied to obtain the maximum fidelity. Figure~\ref{fig:sup_PSB_sim}b has the fidelity at different readout times, $\Delta t$. We obtain a maximum $F_\mathrm{E}=99.3\%$ for a readout time $\Delta t=4$~$\mu$s. 

\section{Leaver arm, electron temperature and tunnel rate of the SEB \label{supp:_Te}}

\begin{figure}[h]
    \centering
    \includegraphics[width=\textwidth]{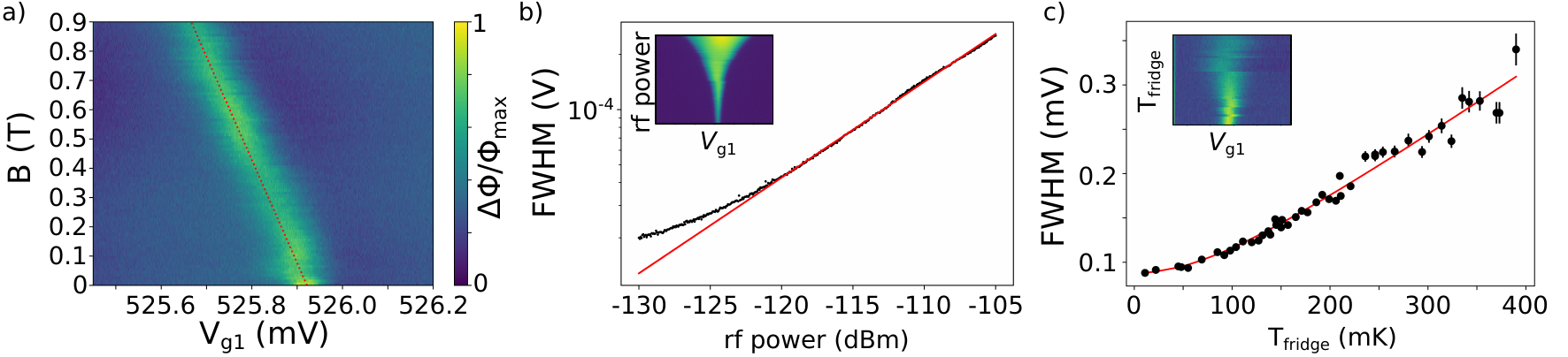}
    \caption{SEB lever arm and electron temperature. a) Magnetospectroscopy of dot to reservoir transition used to extract the SEB lever arm. b) FWHM of the SEB transition as a function of rf power (raw data in inset). Above -120~dBm the deviation of the data (black dots) from the linear fit (red line) indicates the peak is power broadened, thus the temperature dependence of the FWHM in (c) was taken at -130~dBm. From the fit shown in red, we estimate a $T_\text{e}$ of 115$\pm$6 mK.
    }
    \label{fig:supp_E_T}
\end{figure}

To determine the lever arm $\alpha$ of the SEB, we carried out a magnetospectroscopy measurement of the SEB dot to reservoir line close to the ICT of interest, as shown in Fig.~\ref{fig:supp_E_T}a.
From the analysis carried out in \S\ref{supp:_dig}, and due to the linear dependence of the magnetospectroscopy, the transition is temperature broadened and thus the phase response, $\Delta\Phi$, takes the form
\begin{equation}
\label{eq:cosh}
    \Delta \Phi  \propto \frac{1}{\text{cosh}^2\Big(\frac{\alpha(V_\text{g1}-V_\text{g1}^0)}{2k_\text{B}T}\Big)}, 
\end{equation} 
where $V_\text{g1}^0$ is the gate voltage at the center of the peak. The change in $V_\text{g1}^0$ due to the applied magnetic field $B$ is directly related to the SEB's leaver arm, as:
\begin{equation}
    g\mu_B \Delta B = e \alpha \Delta V_\text{g1}^0
\end{equation}
where we take the electron $g$ factor as 2, $\mu_B$ is the Bohr magneton and $e$ is the charge of an electron.
From the fit, we extract an $\alpha$ of 0.40.
To characterise the electron temperature $T_\text{e}$ of the SEB, we measure the FWHM of the SEB as a function of mixing chamber fridge temperature, $T_\text{fridge}$. To ensure the transition is not power broadened, we first measure the transition at varying rf powers at base temperature, as shown in Fig.~\ref{fig:supp_E_T}b. We determine that above -125~dBm the signal becomes power broadened, and thus we take temperature dependence measurements at -130~dBm. To extract the electron temperature $T_\text{e}$ of the SEB, we fit the expression
\begin{equation}
    \text{FWHM} = \frac{3.53k_B}{e\alpha}\sqrt{T_{\text{fridge}}^2+T_e^2}
\end{equation}
where $k_\text{B}$ is the Boltzmann constant and $e$ is the charge of an electron.
From the fit in Fig.~\ref{fig:supp_E_T}c, we estimate an electron temperature of 115$\pm$6~mK.
Since we are thermally broadened, we estimate an upper bound for the tunnel rate $\gamma \le$4.25 GHz, as:
\begin{equation}
    3.53 k_B T_e \ge 2 h \gamma
\end{equation}

\section{Resonator parameter extraction using kinetic inductance changes}
\label{supp:resonator}
\begin{figure}[h]
    \centering
    \includegraphics[width=\textwidth]{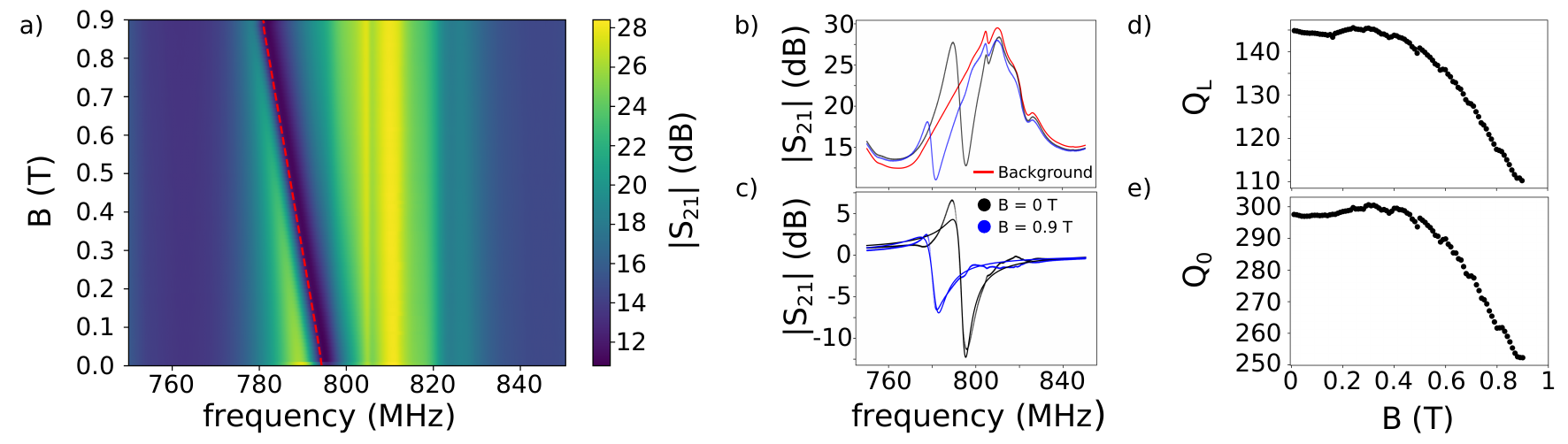}
    \caption{Resonator parameter extraction. a) Magneto-spectroscopy of $|S_{21}|$ as a function of magnetic field $B$, resulting in a change in resonance frequency, as highlighted by the red dashed line. b) $|S_{21}|$ at $B$ fields of 0 T (black) and  0.9 T (blue) with the estimated background signal in red. c) Background subtracted response with corresponding fits, from which the loaded $Q_L$ (d) and internal $Q_0$ (e) quality factors are then extracted as a function of applied field $B$.}
    \label{fig:supp_resonator}
\end{figure}

Due to the presence of uncalibrated standing waves in the reflected RF signal arising from impedance miss-match in the reflectometry setup, magneto-spectroscopy of the reflected signal were taken up to 0.9 T, as shown in Fig~\ref{fig:supp_resonator}a. We assume that the background remains constant while the resonance frequency shifts as a function of $B$ field, allowing us to infer the background amplitude transfer function, as shown in Fig.~\ref{fig:supp_resonator}b.
To account for any asymmetry in the line, we fit a complex external quality factor $\Tilde{Q_\text{e}}$ resulting in the fits shown in Fig.~\ref{fig:supp_resonator}c:
\begin{equation}
    S_{21} = A \Bigg(1 - \frac{2Q_Le^{i\phi}}{|\Tilde{Q_e}|(1 + 2 i Q_L \frac{f - f_\text{r}}{f_\text{r}})}\Bigg)
\end{equation}
The extracted parameters at zero field are a resonance frequency $f_\text{r}$=797 MHz, a loaded quality factor $Q_\text{L}$ of 145, an external quality factor $Q_\text{e}$ of 282, resulting in an intrinsic quality factor $Q_0$ of 298. We get a phase delay $\phi$ of 0.98 and an amplitude $A$ of 17.2 dB, resulting in a matching $\beta$ of 1.05. From the extracted values, we observe that there is no change in $Q_\text{L}$ and $Q_0$ up to 0.4 T (Fig.~\ref{fig:supp_resonator}d,e), above which the resonance appears to deteriorate, potentially due to vortex formation in the superconducting NbN inductor.  

\section{PID controller}\label{supp:PID}

\begin{figure}[h]
    \centering
    \includegraphics[width=\textwidth]{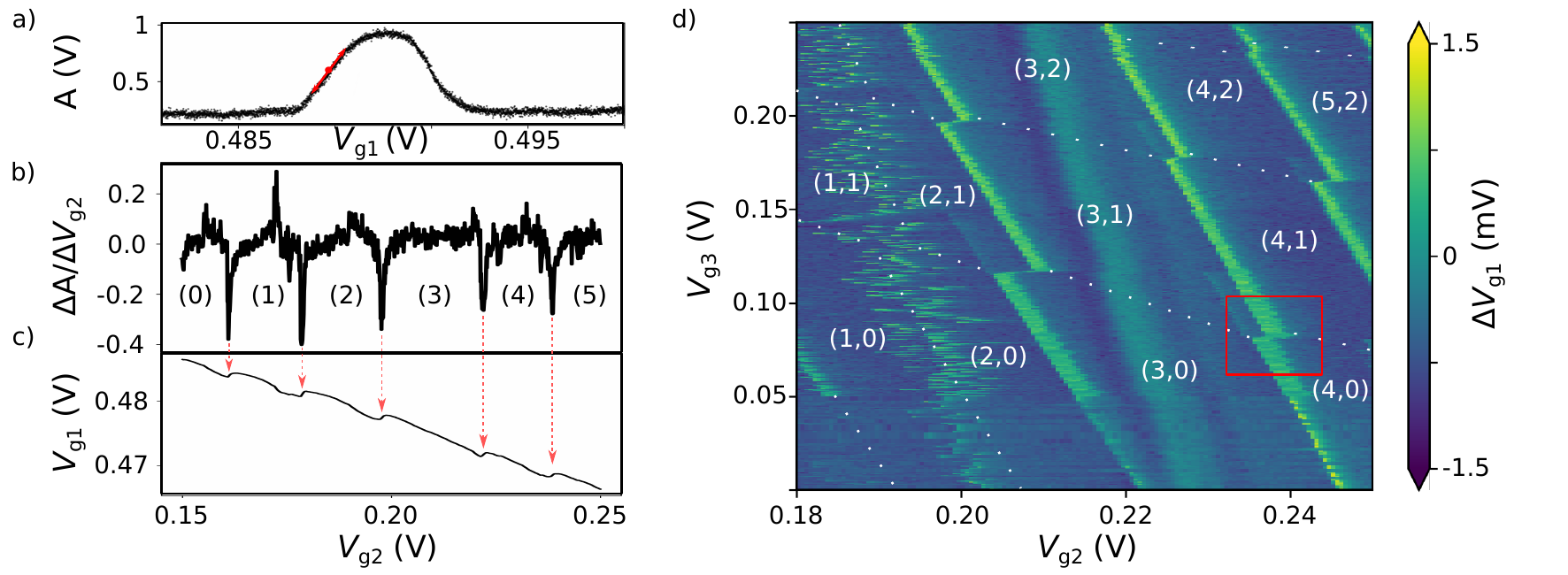}
    \caption{PID charge sensing in the few-electron regime. a) Normalised magnitude of reflected signal of dot to reservoir transition line of the SEB used for charge sensing. The red dot shows the set point used for PID control, with arrows indicating how the signal varies as the peak shifts. b) Change in magnitude response due to a change in V$_\text{g2}$ with the electron occupancy in QD$_2$ in brackets. c) PID response on V$_\text{g1}$ due to the change in signal in (b), highlighted by the red arrows. d) Differential of V$_\text{g1}$ PID response due to a change in V$_\text{g2}$ and V$_\text{g3}$, resulting in the stability diagram in the few-electron regime. We indicate the electron occupancy as well as a red rectangle highlighting the inter-dot charge transition used for single-shot readout.
    } 
    \label{fig:supp_PID}
\end{figure}

A Proportional-Integral-Derivative (PID) controller is a closed loop control system employing feedback to maintain a certain set-point. In this particular implementation, we bias the SEB at the point of its maximum derivative of the magnitude of the reflective signal  $\Delta A$ (red dot in Fig~\ref{fig:supp_PID}a), as this is the most sensitive point. To increase the dynamic range of the PID controller, a large rf power of -100~dBm is applied to broaden the peak, resulting in a FWHM=2~mV. Varying $V_\text{g2}$ (and to a lesser extent $V_\text{g3}$) results in a large change in signal $\Delta A$, as shown in Fig~\ref{fig:supp_PID}b, as it shifts the SEB bias point due to the QDs' cross capacitances. 
Based on the change in signal, the feedback loop compensates $V_{g1}$, as shown in Fig~\ref{fig:supp_PID}c, according to
\begin{equation}\label{eq:PID}
    V_\text{g1}(n+1) = V_\text{g1}(n) + m_i \Delta V_{\text{g}i} + P(n) + I(n) + D(n), \quad [V_\text{g1}^\text{min}, V_\text{g1}^\text{max}].
\end{equation}
To ensure that the PID feedback loop does not go out of range, the controller output is bounded between $V_\text{g1}^\text{min}$ and $V_\text{g1}^\text{max}$, which are user-defined. In Eq.~\eqref{eq:PID}, $V_\text{g1}(n)$ is the voltage on the SEB at step $n$, $m_i$ is the gradient due to the gate capacitance ratio between the SEB and QD$_i$, as estimated in \S\ref{supp:_dig} and $\Delta V_{\text{g}i}$ is the gate voltage step taken on gate $i$, in this implementation $i$=2 or 3.
Then, $P(n)$, $I(n)$ and $D(n)$ are the proportional, integral and differential compensations at step $n$, which are defined as
\begin{align}
    P(n) &= K_\text{P} \Delta A \\
    I(n) &= K_\text{I} \frac{\Delta V_{\text{g}i}}{2} (\Delta A(n) + \Delta A(n-1)) + I(n-1),  \quad [I_\text{min},I_\text{max}]\\
    D(n) &= \frac{2K_\text{D}}{2\tau + \Delta V_{\text{g}i}}  (\Delta A(n) - \Delta A(n-1)) +\frac{2\tau-\Delta V_{\text{g}i}}{2\tau+\Delta V_{\text{g}i}} D(n-1),
\end{align}
where $K_\text{P}$, $K_\text{I}$ and $K_\text{D}$ are the proportional, integral and differential coefficients, which have to be tuned by the user. Here, $\tau$ is the cut-off frequency of a low pass filter used to reduce high frequency noise on the differential term. We use $\tau=1$. The integrator part is limited via a dynamic integrator clamping scheme to ensure that no integration occurs if the signal is already saturated by $P(n)$,
\begin{align}
    I_\text{max} =& \text{max}\big(V_\text{g1}^\text{max} -V_\text{g1}(n+1) +I(n), 0\big) \\
    I_{min} =& \text{min}\big(V_\text{g1}^\text{max} -V_\text{g1}(n+1) +I(n), 0\big).
\end{align}

Once the PID is tuned, we acquire the charge stability diagram shown in Fig.~\ref{fig:supp_PID}d. The PID performs well apart from the region corresponding to $1\leftrightarrow 2$ charge transition of QD$_2$. Nevertheless, we accurately determine the location in gate voltage space of the (3,1)-(4,0) charge transition, as highlighted by the red box, as well as an estimate of the (1,1)-(2,0) transition.

\section{Valley splitting \label{supp:_valley}}

\begin{figure}[h]
    \centering
    \includegraphics[width=\textwidth]{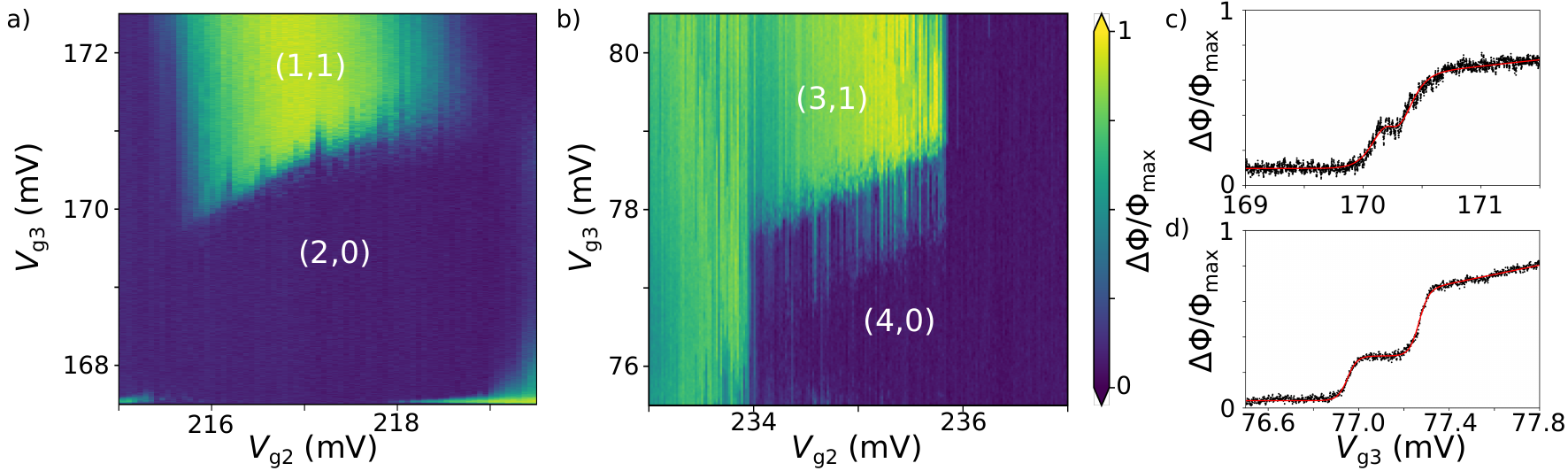}
    \caption{Valley splitting of the (1,1)-(2,0) and (3,1)-(4,0) charge transitions. ICTs of (1,1)-(2,0) (a) and (3,1)-(4,0) (b) in which PSB is observed. Single traces taken while ramping $V_{\text{g3}}$ along SPB for the (1,1)-(2,0) (c) and (3,1)-(4,0) (d). Double Fermi distribution function is then fitted to estimate the valley splittings as 15.6~$\mu$eV and 195.5~$\mu$eV respectively.}
    \label{fig:supp_valley}
\end{figure}

From the estimated voltage space regions from the stability diagram using the PID controller in Fig.~\ref{fig:supp_PID}d, we were able to locate the (1,1)-(2,0) and the (3,1)-(4,0) charge transitions, which exhibit PSB as shown in Fig.\ref{fig:supp_valley}a,b. 
By taking a vertical trace in the region of PSB, two Fermi distribution like functions are observed, as shown in Fig.\ref{fig:supp_valley}c,d. To extract the measurement window $\Delta V$ and the lever arm $\alpha$ of QD$_3$, the following function was fitted to the data:
\begin{equation}
    m_1 \cdot F_1(V_{\text{g3}}) + c_1 + (m_2 \cdot V_{\text{g3}}+c_2) \cdot F_2(V_{\text{g3}})
\end{equation}

Where $m_1$, $m_2$, $c_1$ and $c_2$ are fitting parameters for the linear gradients due to shifts in the charge sensor as a result of ramping $V_{\text{g3}}$. $F_i(V_{\text{g3}})$ is a Fermi distribution function centered at $V^0_i$:
\begin{equation}
    F_i(V_{\text{g3}}) = \frac{1}{\exp\Big(\frac{e\alpha (V_{\text{g3}}-V^0_i)}{k_B T_e}\Big)+1}
\end{equation}
Where $e$ is the charge of an electron, $k_\text{B}$ is the Boltzmann constant  and $T_\text{e}$ is the electron temperature, as determined in section~\ref{supp:_Te}.
From the fits in Fig.~\ref{fig:supp_valley}c,d, we estimate a $\Delta V = (V^0_2-V^0_1)$ of 113 and 374 $\mu$V and an $\alpha$ of 0.139 and 0.525 respectively.
The valley splitting can then be evaluated as $e \alpha \Delta V$, giving an estimate of 15.6~$\mu$eV and 195.5~$\mu$eV respectively.

\section{\label{supp:_dig}Strong charge sensor response}

\begin{figure}[h]
    \centering
    \includegraphics[width=\textwidth]{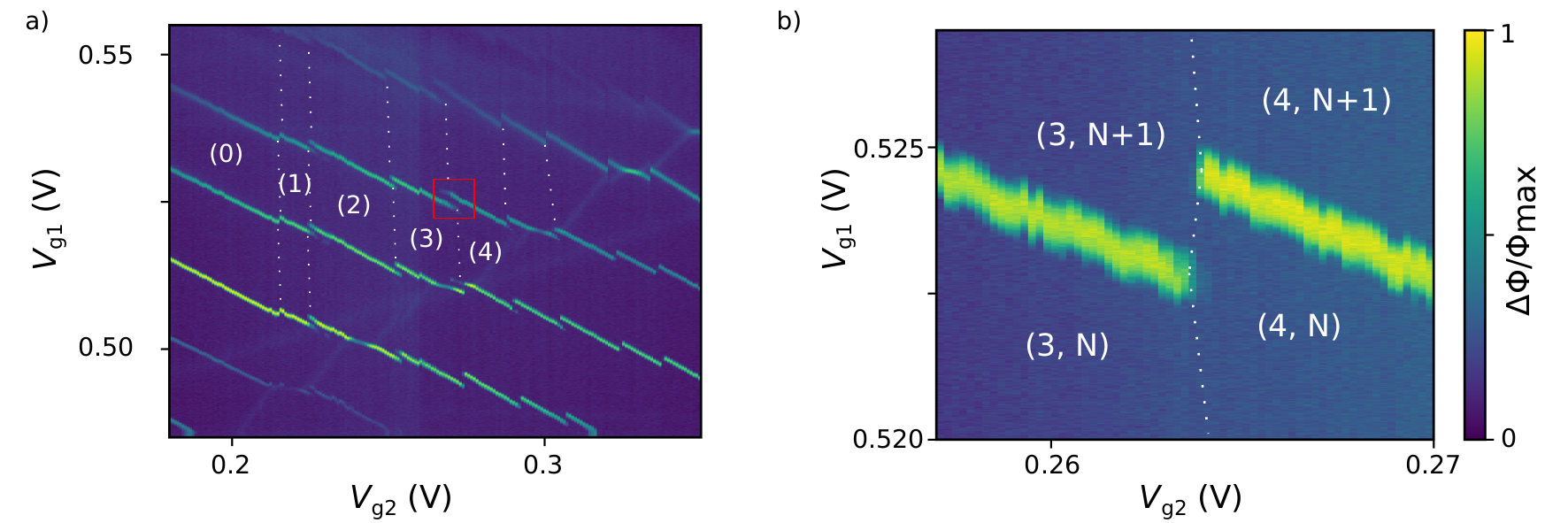}
    \caption{Charge stability diagram $QD_1$ vs $QD_2$. a) Highlighted charge sensing region of interest (red square) in the $V_{\text{g1}}$ vs $V_{\text{g2}}$ gate space. The number numbers indicate the electron occupation in $QD_2$. b) Zoomed in measurement, highlighting the large charge response of the SEB when an electron is loaded on to $QD_2$. The numbers in parenthesis indicate the electron occupation of $QD_2$ and $QD_1$, respectively.}
    \label{fig:supp_g1_g2}
\end{figure}

By operating the QD1 as an SEB, there are multiple dot to reservoir transitions that can be used for charge sensing, as shown in Fig.~\ref{fig:supp_g1_g2}a. Typically, sensing is done with the most intense transition due to the higher signal-to-noise ratio (SNR). However, in our particular system, there is a positively sloped transition line that couples to the SEB, probably due to a defect in the silicon oxide. This line interacts with the SEB close to the $3\leftrightarrow 4$ charge transition of QD$_2$ for the two most intense SEB lines in Fig.~\ref{fig:supp_g1_g2}a. As a result, we operate the SEB at the third most intense line, as highlighted by the red box.
From this data set, and a similar one of $V_\text{g1}$ vs $V_\text{g3}$ (not shown here), the gradients $\frac{\Delta V_\text{g1}}{\Delta V_\text{g2}}$ and $\frac{\Delta V_\text{g1}}{\Delta V_\text{g3}}$ were estimated as -0.237 and -0.034 respectively. 
This was carried out by taking the Hough transform of the threshold data set, as described in ref.~\cite{oakes2020automatic}. These slopes indicate QD$_2$ is seven times more coupled to the SEB than QD$_3$. These slopes were later built in the charge sensing controller of \S\ref{supp:PID}.
From Fig.~\ref{fig:supp_g1_g2}b, we estimate the voltage shift in QD$_1$ when an electron is loaded into QD$_2$. This was done by fitting a cosh$^{-2}$ distribution to the SEB transition for varying $V_\text{g2}$ voltages (Eq.~\ref{eq:cosh}). From the fits, we extract the center and FWHM of the transition line, resulting in a voltage shift $\Delta V_\text{g1}$ of 1.70~mV, which is almost three times the average FWHM of 0.64~mV. We also note that the FWHM is constant as a function of $V_\text{g2}$, indicating that the peak is thermally, rather than lifetime broadened~\cite{Ahmed2018b}.

\section{Singlet triplet model for Pauli spin blockade \label{supp:_ST_PSB}}
For Pauli spin blockade, an analytical model to describe the probability density functions (PDFs) for both a singlet $n_\text{S}$ and a triple $n_\text{T}$ exists~\cite{barthel2009rapid}. In the case of Gaussian noise of equal strength in both quadratures (e.g. cryogenic amplifier noise limited), the singlet probability in the $IQ$ plane reads
\begin{equation}
\label{Eq:nS}
    n_\text{S}(I, Q) = \frac{(1 - <P_\text{T}>)}{2\pi \sigma^2} \cdot \text{exp}\Bigg(-\frac{(I-I^\text{S})^2}{2 \sigma^2} -\frac{(Q-Q^\text{S})^2}{2 \sigma^2}\Bigg),
\end{equation}
where $<P_\text{T}>$ is the probability of a triplet outcome, ($I^\text{S}$, $Q^\text{S}$) is the center of the singlet Gaussian in the $IQ$ plane and $\sigma$ is the standard deviation of the distribution.
For the triplet PDF, there is the added possibility of relaxation to a singlet during the measurement time $\Delta t$, resulting in two extra integral terms
\begin{align}
\label{Eq:nT}
n_\text{T}(I, Q) =& \frac{<P_\text{T}>}{2\pi \sigma^2}\text{exp}\Big(-\frac{\Delta t}{T_1}\Big)  \cdot \text{exp}\Bigg(-\frac{(I-I^\text{T})^2}{2\sigma^2} -\frac{(Q-Q^\text{T})^2}{2\sigma^2}\Bigg)\\
+& \int_{I^\text{S}}^{I^\text{T}}\frac{\Delta t <P_\text{T}>}{T_1(I^\text{T}-I^\text{S})} \text{exp}\Bigg(-\frac{I_x-I^\text{S}}{I^\text{T}-I^\text{S}}\frac{\Delta t}{T_1}\Bigg) \cdot \text{exp}\Bigg(-\frac{(I-I_x)^2}{2\sigma^2}\Bigg)\frac{dI_x}{2\pi\sigma} \nonumber \\
+&\int_{Q^\text{S}}^{Q^\text{T}}\frac{\Delta t <P_\text{T}>}{T_1(Q^\text{T}-Q^\text{S})} \text{exp}\Bigg(-\frac{Q_y-Q^\text{S}}{Q^\text{T}-Q^\text{S}}\frac{\Delta t}{T_1}\Bigg) \cdot \text{exp}\Bigg(-\frac{(Q-Q_y)^2}{2\sigma^2}\Bigg)\frac{dQ_y}{2\pi\sigma}, \nonumber
\end{align}
where the integral terms due to the triplet decaying into a singlet $I_\text{D}$ (second and third addends) have an analytical solution
\begin{align}
    I_\text{D}(V) = \frac{\Delta t <P_\text{T}>}{\sqrt{2\pi}T_1(V^\text{T}-V^\text{S})} \cdot \text{exp}\Big[\frac{\Delta t}{(V^\text{T}-V^\text{S})T_1}(V^\text{S}-V+\frac{\Delta t \sigma^2}{2(V^\text{T}-V^\text{S})T_1})\Big]\\ \cdot \Big[\text{erf}\Big[\frac{\sigma \Delta t}{(V^\text{T}-V^\text{S})T_1\sqrt{2}} + \frac{V^\text{T}-V}{\sigma\sqrt{2}}\Big] - \text{erf}\Big[\frac{\sigma \Delta t}{(V^\text{T}-V^\text{S})T_1\sqrt{2}} + \frac{V^\text{S}-V}{\sigma\sqrt{2}}\Big]\Big].
\end{align}
We use these equations to fit the outcome of the single-shots in the $IQ$ plane (the average signal over $\Delta t$). We extract the centers of the two distributions. We then project the data along the axis that connects both centers, to thus reduce the problem to one dimension, as in Barthel et al.~\cite{barthel2009rapid} and hence reduce the number of fitting parameters to $<P_\text{T}>$, $V_T$ and $\sigma$. From the fitted parameters, we determine the singlet $F_\text{S}$ and triplet $F_\text{T}$ fidelities as a function of threshold voltage $V_\text{T}$ above which we call the shot a triplet,

\begin{align}
    F_\text{S} =& 1 - \int_{V_\text{T}}^{\infty} \frac{n_\text{S}(V)}{1 - <P_\text{T}>} dV =  1-\frac{1}{2}\Big[1+\text{erf}\Big(\frac{V_\text{S}-V}{\sqrt{2}\sigma}\Big)\Big] \\
    F_\text{T} =& 1 - \int_{-\infty}^{V_\text{T}} \frac{n_\text{T}(V)}{<P_\text{T}>} dV =  1-\frac{1}{2}\text{exp}(-\frac{\Delta t}{T_1})\Big[1-\text{erf}\Big(\frac{V_\text{T}-V}{\sqrt{2}\sigma}\Big)\Big] -\int_{-\infty}^{V_\text{T}} I_\text{D}(V) dV.
\end{align}

Here the integral for $F_\text{T}$ has no analytical solution, and thus needs to be computed numerically. The visibility $V_\text{E}$ and the average spin measurement fidelity, $F_\text{M}$, are then determined as
\begin{equation}
    V_\text{E} = \text{max}\Big(F_\text{S}(V_\text{T}) + F_\text{T}(V_\text{T}) - 1\Big) \quad \quad V_\text{M} = \text{max}\Big((1-<P_\text{T}>)F_\text{S}(V_\text{T}) + <P_\text{T}>F_\text{T}(V_\text{T})\Big).
\end{equation}

\section{Signal to noise ratio and ideal fidelity} \label{supp_SNR}

\begin{figure}[h]
    \centering
    \includegraphics[width=\textwidth]{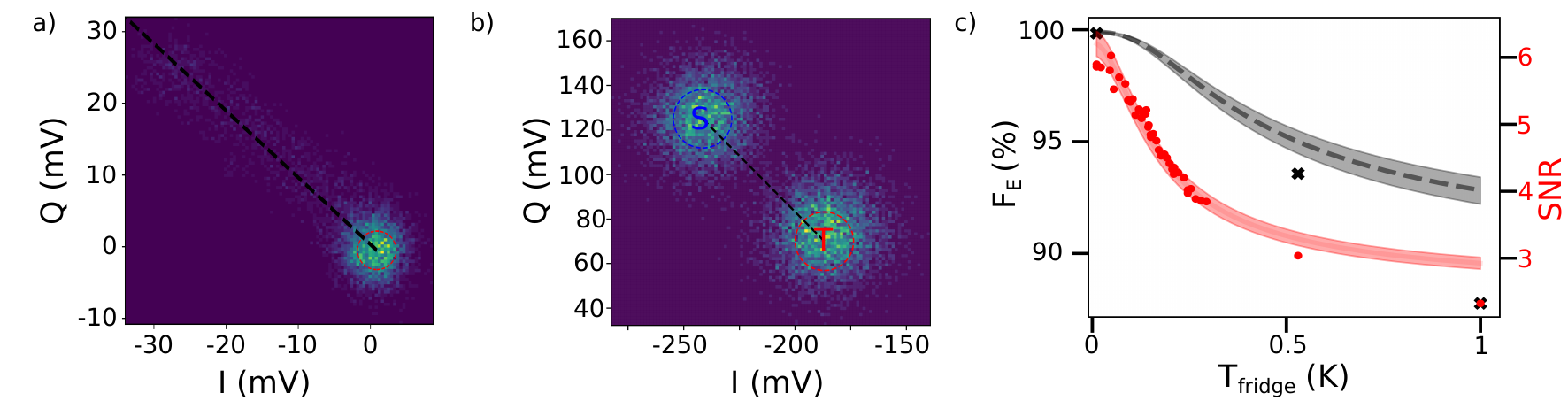}
    \caption{SNR extraction for a) DTR transition on sensing QD and b) single shot data. c) SNR data-points as a function of temperature, with best fit (red line). From the fitted SNR data, the electrical fidelity was estimated for a measurement time $\Delta t$ of 5.6~$\mu$s (black line). Black crosses are the measured electrical fidelity.}
    \label{fig:sup_SNR}
\end{figure}

To extract the SNR for the SEB at various temperatures, each trace from the insert in Fig.~\ref{fig:supp_E_T}c was plotted in the IQ plane, as shown in Fig~\ref{fig:sup_SNR}a. Since we subtract the average background for the IQ data, the background noise forms a 2D Gaussian distribution centred at the origin. The circular shape shows the system is cryogenic amplifier noise limited. The standard deviation of the Gaussian distribution (red circle) represents the noise of the system, while the signal is measured as the distance between the maximum value, obtained by fitting Eq~\ref{eq:cosh} to the I and Q data and the centre of the noise, depicted by the black dashed line. A similar procedure is carried out for the single-shot data, after averaging the signal over the measurement time $\Delta t$, as shown in Fig~\ref{fig:sup_SNR}b. The main difference is that two peaks appear, one due to the singlets and the other due to the triplets outcomes, which we fit according to Eq.~\eqref{Eq:nS} and~\eqref{Eq:nT}, respectively. The noise is still calculated as the fitted average standard deviation, while the signal is the distance between the two peaks. 
Since the two methods adopted to measure the SNR were carried out at two different rf powers and on different days, the SNR values extracted at base temperature were used to calibrate the two data-sets, resulting in the red data-points on Fig~\ref{supp_SNR}c. While the SEB peak shift remains greater than its FWHM, the SNR has the following temperature dependence
\begin{equation}
    \text{SNR} \propto \frac{1}{\sqrt{T_{\text{fridge}}^2+T_\text{e}^2}},
\end{equation}
where the value extracted in \S\ref{supp:_Te} was used for $T_\text{e}$, resulting in the fitted red curve. At 1 K the fitted SNR is larger than the measured one, suggesting that other factors may be deteriorating the overall SNR. From the SNR, one can then calculate the electrical fidelity $F_\text{E}$ (assuming an infinite $T_1$), resulting in the blue dotted line:
\begin{equation}
    F_\text{E} = \frac{1}{2}\Bigg(1+\text{erf}\Bigg(\frac{\text{SNR}}{2\sqrt{2}}\Bigg)\Bigg)
\end{equation}

